\documentclass[%
 reprint,
superscriptaddress,
 amsmath,amssymb,
 aps,
pra,
]{revtex4-1}

\usepackage{amsthm,amssymb}
\usepackage{amsmath}
\usepackage{graphicx}% Include figure files
\usepackage{dcolumn}% Align table columns on decimal point
\usepackage{bm}% bold math
\usepackage{multirow} % multirow
\usepackage{mathdots}
\usepackage{enumerate}
\usepackage{algorithm}
\usepackage{algorithmicx}
\usepackage{algpseudocode}
\usepackage{textcomp}
\usepackage{color}

\usepackage{epstopdf}
\usepackage{epsfig}
\usepackage{array}
\usepackage{booktabs}
\usepackage{subfigure}%
\usepackage{setspace}%
\usepackage{amsmath}%
\usepackage{tikz}%
\usepackage{hyperref}
\usepackage{multirow}
\usepackage{changes}

\usetikzlibrary{shapes,arrows}

\renewcommand{\algorithmicrequire}{\textbf{Input:}}
\renewcommand{\algorithmicensure}{\textbf{Output:}}
\theoremstyle{remark}

\renewcommand{\algorithmicrequire}{\textbf{Input:}} % Use Input in the format of Algorithm
\renewcommand{\algorithmicensure}{\textbf{Output:}} % Use Output in the format of Algorithm

\begin{document}

\makeatletter
\newcommand{\ud}{\mathrm{d}}
\newcommand{\rmnum}[1]{\romannumeral #1}
\newcommand{\polylog}{\mathrm{polylog}}
\newcommand{\ket}[1]{|{#1}\rangle}
\newcommand{\bra}[1]{\langle{#1}|}
\newcommand{\inn}[2]{\langle{#1}|#2\rangle}
\newcommand{\Rmnum}[1]{\expandafter\@slowromancap\romannumeral #1@}
\newcommand{\udots}{\mathinner{\mskip1mu\raise1pt\vbox{\kern7pt\hbox{.}}
        \mskip2mu\raise4pt\hbox{.}\mskip2mu\raise7pt\hbox{.}\mskip1mu}}
\makeatother

\preprint{APS/123-QED}

\title{Multilevel leapfrogging initialization strategy for quantum approximate optimization algorithm}

\author{Xiao-Hui Ni}
\affiliation{State Key Laboratory of Networking and Switching Technology, Beijing University of Posts and Telecommunications, Beijing, 100876, China}
%\affiliation{State Key Laboratory of Cryptology, P.O. Box 5159, Beijing, 100878, China}
\author{Bin-Bin Cai}
\affiliation{College of Computer and Cyber Security, Fujian Normal University, FuZhou ,350117, China}
\author{Hai-Ling Liu}
\affiliation{State Key Laboratory of Networking and Switching Technology, Beijing University of Posts and Telecommunications, Beijing, 100876, China}
\author{Su-Juan Qin}
\email{qsujuan@bupt.edu.cn}
\affiliation{State Key Laboratory of Networking and Switching Technology, Beijing University of Posts and Telecommunications, Beijing, 100876, China}
\author{Fei Gao}
\email{gaof@bupt.edu.cn}
\affiliation{State Key Laboratory of Networking and Switching Technology, Beijing University of Posts and Telecommunications, Beijing, 100876, China}
\author{Qiao-Yan Wen}
%\email{wqy@bupt.edu.cn}
\affiliation{State Key Laboratory of Networking and Switching Technology, Beijing University of Posts and Telecommunications, Beijing, 100876, China}

\date{\today}
\begin{abstract}
	Recently, Zhou et al. have proposed a novel Interpolation-based (INTERP) strategy to generate the initial parameters for the Parameterized Quantum Circuit (PQC) in Quantum Approximate Optimization Algorithm (QAOA). INTERP produces the guess of the initial parameters at level $i+1$ by applying linear interpolation to the optimized parameters at level $i$, achieving better performance than random initialization (RI). Nevertheless, INTERP consumes extensive running costs for deep QAOA because it necessitates optimization at each level of the PQC. To address this problem, a Multilevel Leapfrogging Interpolation (MLI) strategy is proposed. MLI can produce the guess of the initial parameters from level $i+1$ to $i+l$ ($l>1$) at level $i$, omitting the optimization rounds from level $i+1$ to $(i+l-1)$. The final result is that MLI executes optimization at few levels rather than each level, and this operation is referred to as Multilevel Leapfrogging optimization (M-Leap). The performance of MLI is investigated on the Maxcut problem. Compared with INTERP, MLI reduces most optimization rounds. Remarkably, the simulation results demonstrate that MLI can achieve the same quasi-optima as INTERP while consuming only 1/2 of the running costs required by INTERP. In addition, for MLI, where there is no RI except for level $1$, the greedy-MLI strategy is presented. The simulation results suggest that greedy-MLI has better stability (i.e., a higher average approximation ratio) than INTERP and MLI beyond obtaining the same quasi-optima as INTERP. According to the efficiency of finding the quasi-optima, the idea of M-Leap might be extended to other training tasks, especially those requiring numerous optimizations, such as training adaptive quantum circuits.
\end{abstract}
\pacs{Valid PACS appear here}
\maketitle

\section{Introduction}
%\justify
With advancements in quantum computing technology, many quantum algorithms have demonstrated significant speed advantages in solving certain problems, such as unstructured database searching \cite{grover}, dimensionality reduction \cite{qpca,pan_1,Yu3}, solving equations \cite{hhl,wan_1} and ridge regression \cite{Yu4}. However, noise and qubit limitations prevent serious implementations of the aforementioned quantum algorithms in the current Noisy Intermediate-Scale Quantum (NISQ) devices \cite{NISQ,VQA_review2}. As such, hybrid quantum-classical algorithms \cite{adaptive-vqe, QAS, qubit-adaptive-vqe, Huang,autoencoder_enhance,vqsd,nonlinear,poisson,VQNHE,Wu_2023} are proposed to fully exploit the power of NISQ devices.
\medskip

Quantum Approximate Optimization Algorithm (QAOA) \cite{automatic,alternating_QAOA_ansatz,adaptive-qaoa,phase_transition,performance_QAOA,benchmarking} is a class of hybrid quantum-classical algorithms, presented by Farhi et al. \cite{qaoa} to tackle combinatorial optimization problems such as $k$-vertex cover \cite{minimum_vertex_cover} and exact cover \cite{exact_cover}. In QAOA, the solution of the combinatorial optimization problem tends to be encoded as the ground state of the target Hamiltonian $H_{C}$. The ground state can be approximately obtained by the following steps \cite{VQA_review1}. (i) Build a Parameterized Quantum Circuit (PQC) by utilizing $p$-level QAOA ansatz \cite{TQA}, where one level consists of two unitaries: one is generated by $H_{C}$, the other is generated by the initial Hamiltonian $H_{B}$ \cite{qaoa}. One level has two variational parameters $\gamma_{i}$ and $\beta_{i}$, and there are $2p$ variational QAOA parameters at level $p$, where $i=1,2,\cdots,p$ and $p$ is the level depth. (ii) Initialize $2p$ QAOA parameters for the quantum circuit. (iii) The quantum computer computes the expectation value of the output state and delivers the relevant information to the classical computer. (iv) The classical computer updates the parameters by the classical optimizer and delivers new parameters to the quantum computer. (v) Repeat (iii)-(iv) to maximize (or minimize) the expectation values until meeting a termination condition, and the quantum circuit outputs an estimate of the solution to the problem. 
\medskip

\begin{figure}[!t]
	\centering
	\begin{minipage}[c]{.8\linewidth}
		\centering
		\includegraphics[width=0.9\textwidth]{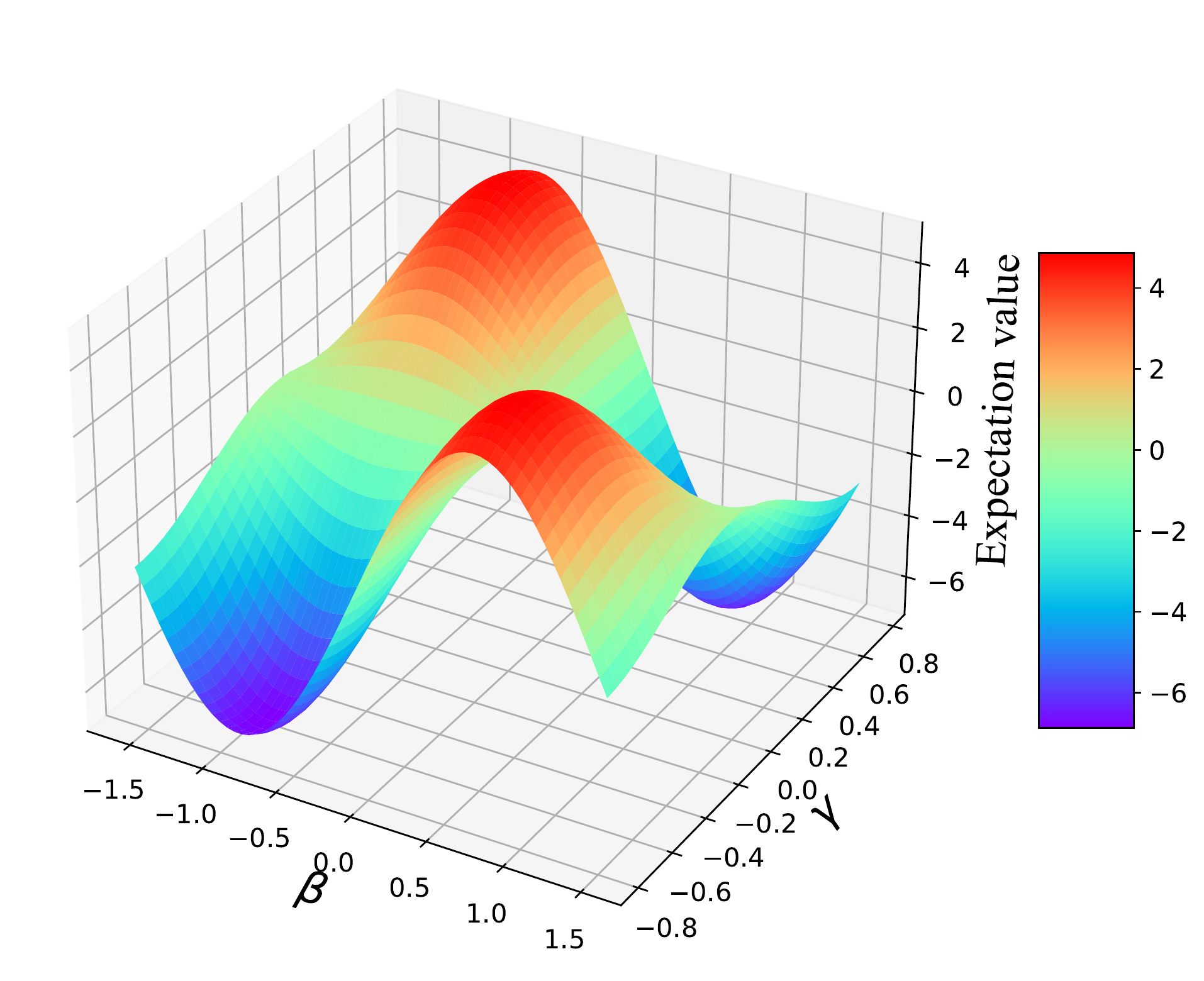}
	\end{minipage}\\[3mm]
	\caption{Energy landscape of QAOA objective function for Maxcut on a 3-regular graph with vertex number $n=10$. There are two variational parameters $\gamma$ and $\beta$ when $p=1$, where we restrict $\gamma \in \left [ \frac{-\pi }{4} ,\frac{\pi }{4} \right ) $ and $\beta \in \left [ \frac{-\pi }{2} ,\frac{\pi }{2} \right ) $ by symmetries as mentioned in Ref.~\cite{INTERP}. In this landscape, the redder region corresponds to high-quality parameters, and it is clear that there are many low-quality parameters.}
	\label{landscape}
\end{figure}

To find the quasi-optimal solution (i.e., nearly global optima), the classical outer loop optimization of QAOA tends to consume abundant costs to search for the quasi-optimal parameters of the PQC \cite{tree-QAOA}, which becomes a performance bottleneck of QAOA \cite{TQA}. Fortunately, the costs can be significantly reduced when QAOA starts from initial parameters that are in the vicinity of quasi-optimal solutions \cite{benchmarking_1, VQA_review1, bilinear}. However, getting such high-quality initial parameters is intractable in the highly non-convex space \cite{INTERP} which is filled with low-quality local optima. Thus, searching for heuristic parameter initialization strategies to effectively tackle this problem becomes a pressing concern \cite{TQA}. To visualize the properties of the space, we give \textbf{Figure~\ref{landscape}} consisting of a three-dimensional graph to depict an energy landscape of the QAOA objective function for Maxcut.

\medskip
Recent studies of heuristic parameter initialization strategies mainly include two aspects. On the one hand, some studies predict QAOA parameters by taking advantage of machine learning techniques \cite{tree-QAOA,mc1,mc2,mc3,mc4}. They have the potential to find good initial parameters, which can accelerate the convergence of QAOA and improve its performance on various optimization problems. However, the existing machine learning models have defects in generalization performance. In other words, numerous models are required for different problem sizes or QAOA depths \cite{PPN}. On the other hand, there are some non-machine learning methods that initialize QAOA parameters based on the properties of parameters \cite{bilinear, INTERP, TQA,fixing_strategy,leapfrogging,angle_conjecture,param_concentration}, such as reusing parameters from similar graphs \cite{leapfrogging} and using the fixed angle conjecture \cite{angle_conjecture}. Building on the observed patterns of the optimal parameters, Zhou et al. \cite{INTERP} proposed an Interpolation-based (INTERP) strategy. INTERP generates the guess of the initial QAOA parameters for level $i+1$ by applying interpolation on the optimized parameters at level $i$. INTERP can find the quasi-optimal solutions in $O$[poly($p$)] time, while random initialization (RI) requires $2^{O(p)}$ optimization runs to achieve similar performance. Nevertheless, INTERP consumes numerous costs (e.g., the number of iterations) in an optimization run because it requires executing one round of optimization at each level of the PQC. Generally, the expectation function requires more iterations to reach convergence as level $i$ increases, consuming more running costs to find the quasi-optimal parameters. Besides, there are other strategies \cite{bilinear,fixing_strategy,layer_vqe, recursive} confronting the same problem.

\medskip
To reduce the costs in an optimization run for deeper QAOA, we present a Multilevel~Leapfrogging~Interpolation (MLI) strategy. MLI concurrently produces the guesses of the initial QAOA parameters from level $i+1$ to $i+l$ at level $i$, where the depth step $l>1$, cutting down on all the optimization rounds from level $i+1$ to $(i+l-1)$. The final result is that MLI executes optimization at few levels rather than each level, and we refer to this operation as Multilevel~Leapfrogging~optimization (M-Leap). We then benchmark the performance of MLI on the Maxcut problem. Compared with INTERP, MLI cuts down on most optimization rounds for the classical outer learning loop. Notably, the simulation results demonstrate that MLI can achieve the same quasi-optimal solutions as INTERP while consuming only 1/2 of the running costs required by INTERP. In addition, for MLI, where there is no RI except for $i=1$, we present greedy-MLI. From level $i=1$ to $i=p$, greedy-MLI starts from multiple sets of the initial parameters and explores the parameter space by parallel optimization. After each optimization round, greedy-MLI preserves all sets of optimized parameters corresponding to the best local maximum in multiple optimization results. Notably, multiple sets of initial QAOA parameters for the new level are generated by respectively applying the interpolation technique on these retained parameters, instead of generating multiple sets of initial parameters by reusing parameters and RI as done in the parameter fixing strategy \cite{fixing_strategy}. The simulation results show that greedy-MLI can get not only the same quasi-optimal solution as INTERP but also a higher average performance than MLI, INTERP, and RI. 

\medskip

Though this paper specifically considers QAOA, the idea of M-Leap might be extended to other training tasks, particularly those demanding extensive optimization efforts to find optimal parameters, such as quantum circuit design \cite{adaptive-vqe,QAS,qubit-adaptive-vqe,adaptive-qaoa,adaptive-vqe1}, quantum circuit learning \cite{QCL,QCL2} and quantum machine learning \cite{mc1,mc2,mc3}, resulting in faster and more resource-efficient quantum computations across different application domains. In addition, MLI and greedy-MLI might be used as subroutines or preprocessing tools in other quantum optimization algorithms \cite{bilinear,leapfrogging,angle_conjecture}, accelerating the exploration of solution spaces and increasing the likelihood of finding quasi-optima for complex problems, thus enhancing the practicality of quantum optimization techniques. The paper is organized as follows. Section~\ref{sec:rev} briefly reviews the Maxcut problem, QAOA, and INTERP. In Section~\ref{sec:our_job}, we propose two heuristic strategies for initializing optimization, that is, MLI and greedy-MLI. Subsequently, we compare the performance of three strategies in Section~\ref{sec:VS}. Finally, a short conclusion and discussion are given in Section~\ref{conclusion}.

\section{Preliminaries}\label{sec:rev}

In this section, we review some relevant preliminaries to help readers better understand our work.

\subsection{Review of Maxcut}\label{Maxcut}

Maxcut is defined in graph $G=(V,E)$, where $V = \left \{ {0,\cdots,n-1} \right \}$ is the set of vertices, $ E = \left \{ (\left ( u,v \right ), w_{u,v}) \right \} $ is the set of edges, $w_{u,v}$ is the weight of edge $(u,v)$ and $w_{u,v}=1$ in this paper. Maxcut aims at maximizing the number of cut edges by dividing $V$ into vertex subsets $S$ and $T$ that do not intersect each other \cite{maxcut,perform1}. 

\medskip
In general, $x_{u} =1 $ if the vertex $u$ is in the subset $S$, otherwise, $x_{u} = 0$, where $u = 0,1,\cdots,n-1$. It is straightforward that every way of vertex partition corresponds to a unique bit string $x$, and there are $2^{n}$ forms for $x$ in total. The number of cut edges pluses one if vertices $u$ and $v$ of edge $e\in E$ are in distinct subsets of vertex, and the goal for Maxcut is to maximize the cost function
\begin{equation}
	C(x) = \sum_{(u,v) \in E} x_{u}+x_{v}-2x_{u}x_{v}.\label{cost_ function}
\end{equation} 

\subsection{Review of QAOA}\label{QAOA}

QAOA is motivated by adiabatic quantum computation \cite{adiabatic}. The basic idea behind QAOA is to start from the ground state of the initial Hamiltonian $H_{B}$ and gradually evolve to the ground state of the target Hamiltonian $H_{C}$ through $p$-level QAOA ansatz \cite{qaoa}. The initial Hamiltonian is conventionally chosen as $H_{B}= -\sum_{j=1}^{n} \sigma _{j}^{x}$ whose ground state $|s\rangle = |+\rangle ^{\otimes n}$ can be effective to prepare, where $\sigma_{j}^{x}$ refers that applying Pauli-X to the $j$-th qubit. To encode the solution of Maxcut into the ground state of $H_{C}$, we convert the cost function $-C(x)$ to the target Hamiltonian
\begin{equation}
	H_{C}=\sum_{(u,v) \in E}\frac{\sigma_{u}^{z}\sigma_{v}^{z} - I }{2}\label{target_ham}
\end{equation} 
by transforming each binary variable $x_{u}$ to a quantum spin $\frac{I-\sigma_{u}^{z} }{2} $, where $\sigma_{u}^{z}$ refers that applying Pauli-Z to the $u$-th qubit.

The ground state of $H_{C}$ can be approximately obtained by alternately applying unitaries $e^{-i\gamma _{i} H_{C}}$ and $e^{-i\beta _{i}  H_{B} }$ on the initial quantum state $|s\rangle$. The output state of the parameterized quantum circuit is 
\begin{equation}
	|\boldsymbol{\gamma_{p}},\boldsymbol{\beta_{p}}\rangle  =e^{-i\beta _{p}  H_{B} }e^{-i\gamma _{p}  H_{C} }\cdots e^{-i\beta _{1}  H_{B} }e^{-i\gamma _{1}  H_{C} } |s\rangle, \label{quantum_state}
\end{equation} 
where $\boldsymbol{\gamma_{p}} = (\gamma_{1},\cdots,\gamma_{p})$ and $\boldsymbol{\beta_{p}} = (\beta_{1},\cdots,\beta_{p})$ are $2p$ variational QAOA parameters. We then define the expectation value $F(\boldsymbol{\gamma_{p}},\boldsymbol{\beta_{p}})$ of $H_{C}$ in this variational quantum state, where
\begin{equation}
	F({\boldsymbol{\gamma_{p}}}  ,{\boldsymbol{\beta_{p}}})	= - \langle {\boldsymbol{\gamma_{p}}},{\boldsymbol{\beta_{p}}} |H_{C}|{\boldsymbol{\gamma_{p}}}  ,{\boldsymbol{\beta_{p}}}\rangle ,\label{expectation}
\end{equation} 
which is done by repeated measurements of the quantum system using the computational basis and our goal is to search for the global optimal parameters 
\begin{equation}
	(\boldsymbol{\gamma_{p}}^{opt},\boldsymbol{\beta_{p}}^{opt}) = \arg  \underset{(\boldsymbol{\gamma_{p}},\boldsymbol{\beta_{p}})}{\max} F(\boldsymbol{\gamma_{p}},\boldsymbol{\beta_{p}}).
\end{equation} 

This search is typically done by starting with some guesses of the initial parameters and performing optimization until reaching the iteration stop condition \cite{INTERP}, such as the maximum number of iterations or $\Delta_{1}$ and $\Delta_{2}$ are lower than $\Delta$ \cite{multistart}, where
\begin{equation}
	\begin{aligned}
		\Delta_{1} = \lvert  F({\boldsymbol{\gamma_{p}^{iter+1}}}  ,{\boldsymbol{\beta_{p}^{iter+1}}})  - F({\boldsymbol{\gamma_{p}^{iter}}}  ,{\boldsymbol{\beta_{p}^{iter}}}) \rvert,\\
		\Delta_{2} = \lvert F({\boldsymbol{\gamma_{p}^{iter}}}  ,{\boldsymbol{\beta_{p}^{iter}}})-F({\boldsymbol{\gamma_{p}^{iter-1}}}  ,{\boldsymbol{\beta_{p}^{iter-1}}}) \rvert,
	\end{aligned} \label{stop_condition}
\end{equation}
where $\Delta$ is the convergence tolerance on function values and the superscript $iter$ means the current iteration, and $F({\boldsymbol{\gamma_{p}^{iter}}} ,{\boldsymbol{\beta_{p}^{iter}}})$ is the related expectation function value. Generally speaking, the lower the $\Delta$, the more iterations may be consumed. To quantify the quality of the QAOA solution, we introduce the approximation ratio (AR)
\begin{equation}
	r = \frac{F({\boldsymbol{\gamma_{p}}}  ,{\boldsymbol{\beta_{p}}})}{F_{max} }, \label{ratio}
\end{equation} 
where $F_{max}$ is the negative of the ground state energy of $H_{C}$. Here, $F_{max}$ equals the maximum cut value of the graph, which can be obtained by brute force search for small graph instances. However, in scenarios where obtaining an exact $F_{max}$ for large graph instances within a reasonable time frame becomes impractical, we can execute multiple runs of the Goemans-Williamson algorithm \cite{GW} and take the maximal value, which can serve as the approximate value of $F_{max}$ to roughly calculate $r$.
The AR reflects how close the solution given by QAOA is to the true solution, $r \le 1$, with the value of 1 nearer to the true solution \cite{bilinear}. In this paper, we obtain $F_{max}$ by brute force search to precisely calculate $r$.

\subsection{Review of INTERP}\label{review_interp}
\begin{figure*}[htbp]
	\centering
	\setlength{\abovecaptionskip}{0.cm}
	\subfigure{
		\includegraphics[width=0.45\textwidth]{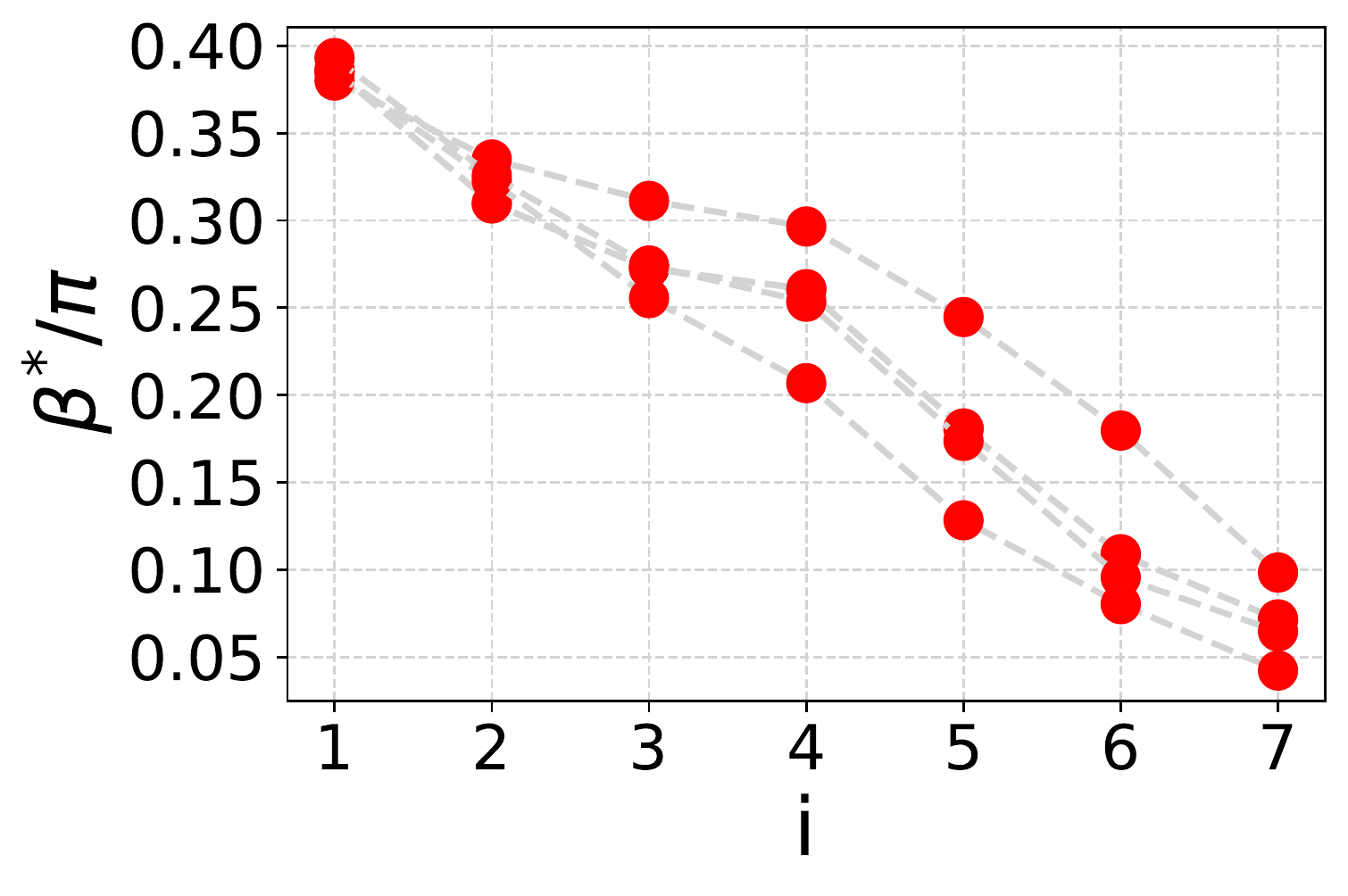}
		\includegraphics[width=0.45\textwidth]{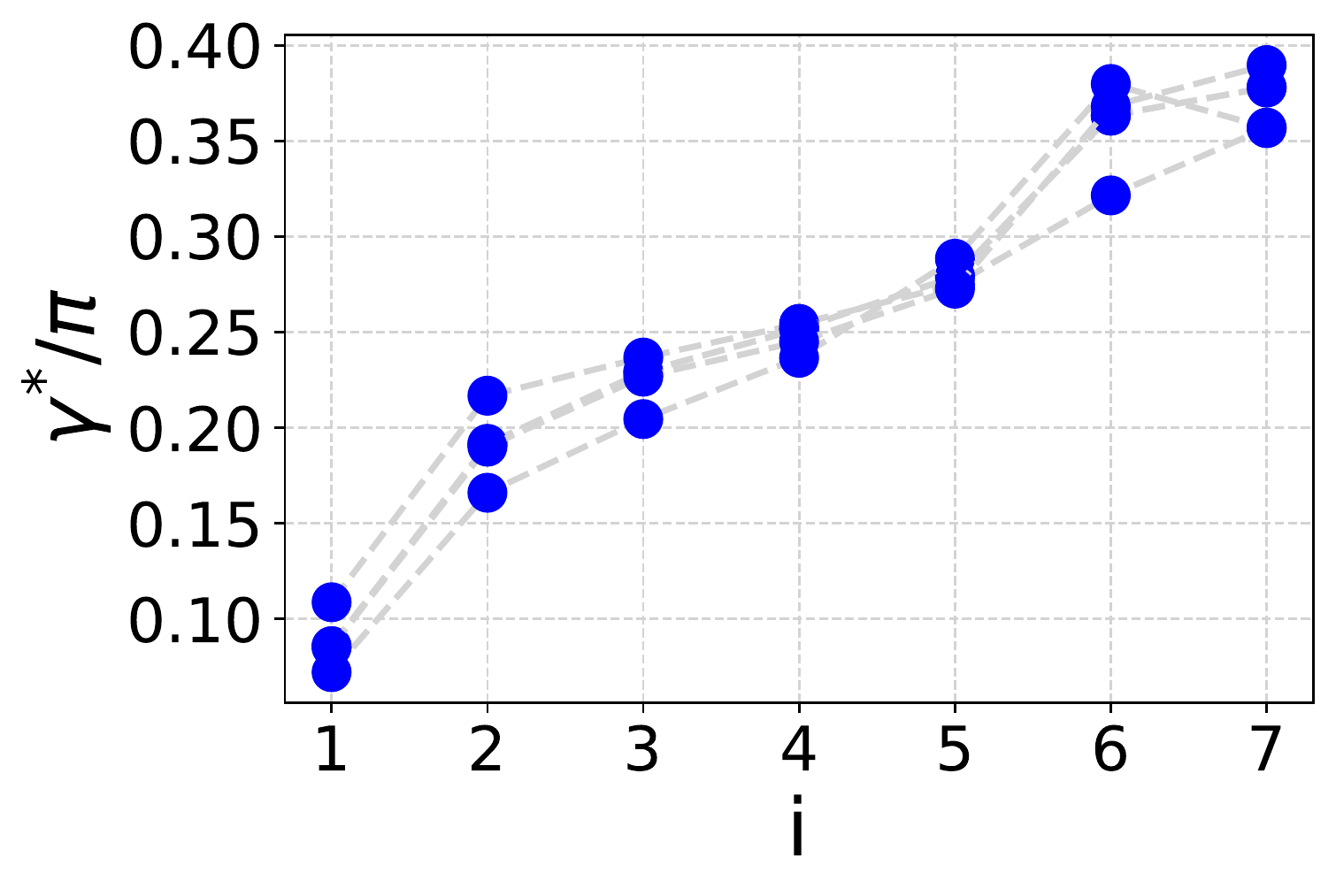}
	}\caption{ The smooth parameter pattern is visualized by plotting the optimal QAOA parameters of four 3-regular graph instances with 8-vertex for $p = 7$. Each dashed line connects parameters for one particular graph instance. For each instance, we execute 1000 RI and save the optimal parameter, corresponding to the optimal expectation function value at $p = 7$, following the approach outlined in Ref.~\cite{INTERP}. Here, the X-axis means the $i$-th level. The Y-axis represents the value of the $i$-th optimal QAOA parameter.}
	\label{param_opt}
\end{figure*}

For the Maxcut problem on numerous 3-regular graphs, Zhou et al. \cite{INTERP} investigated the optimal QAOA parameters by extensive searches in the entire parameter space, and they presented INTERP building on the observed patterns in optimal parameters, where the optimal $\gamma_{i}^{*}$ tends to increase smoothly with $i = 1,2,\cdots,p$, whereas the optimal $\beta_{i}^{*}$ tends to decrease smoothly, as shown in \textbf{Figure~\ref{param_opt}}. The smooth pattern means that the variational QAOA parameters change gradually and continuously as the QAOA depth increases, without sharp jumps or discontinuities \cite{recursive}. The smoothness property can be visually inspected by plotting the changes of QAOA angles as a function of $p$ and observing whether the curve appears continuous and smooth.

\medskip

In the following, we provide a detailed description of INTERP. The strategy works as follows: INTERP optimizes QAOA parameters starting from level $i=1$ and increments $i$ by one after optimizing. For $i = 1$, the initial parameters are randomly generated and then  optimized. For $i\ge 2$, INTERP produces the guess of initial parameters for level $i$ by applying linear interpolation to the optimized parameters at level $i-1$, then optimizes 2$i$ QAOA parameters and increments $i$ by one after optimization. INTERP repeats the above process until it gets the optimized parameters at the target level $p$. INTERP has the capability to find quasi-optimal $p$-level QAOA parameters with a computational complexity of $O$[poly($p$)], and additional optimization runs are necessary to identify quasi-optimal parameters in the highly non-convex parameter space as the level depth $p$ increases. Broadly speaking, as the level $i$ increases within the range of $(1,2,\cdots,p)$, INTERP may necessitate more iterations to converge to a higher expectation value at each level $i$, leading to numerous costs in an optimization run. All in all, executing optimization procedures at each level introduces numerous computational overheads, particularly when dealing with deeper level depths.

\begin{figure}[ht]
	\centering
	\setlength{\abovecaptionskip}{0.cm}
	\subfigure{
		\includegraphics[width=0.45\textwidth]{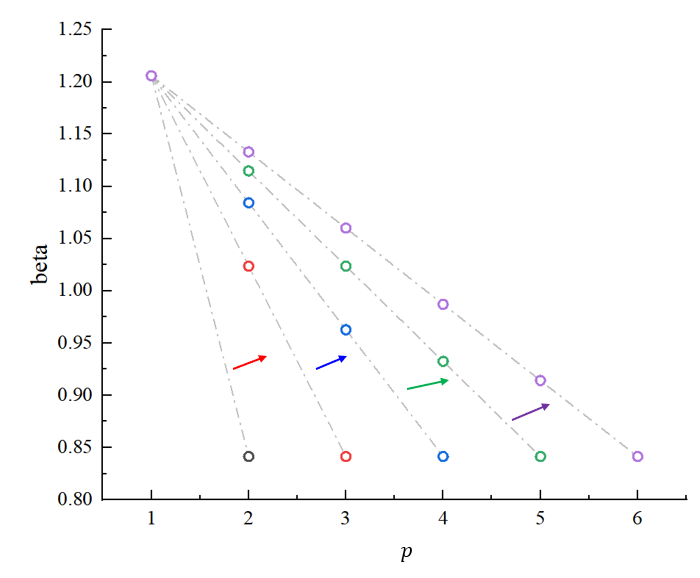}
	}
	\caption{ A graphic example of consecutive interpolations. From $p$ ($\ge 2$) to $p+l$, it corresponds to consecutive interpolations when $l>1$. In each round of interpolation, the first and last parameters are unaltered, but the initial parameter of the $t$-th point at level $p+1$ is altered by the $(t-1)$-th and $t$-th parameters at level $p$, where $t = 2,\cdots,p-1$. Specially, the initial parameters produced by interpolation for $p = 2$ are $\boldsymbol{\gamma_{2}^{0}} = (\gamma_{1}^{*}, \gamma_{1}^{*})$ and $\boldsymbol{\beta_{2}^{0}} = (\beta_{1}^{*}, \beta_{1}^{*})$, where $\gamma_{1}^{*}$ ($\beta_{1}^{*}$) is the optimized parameter at $p = 1$. In the example, we generate the initial parameters for level $p=6$ by applying four consecutive interpolations to the optimized parameters at level $p=2$, where the arrows represent the order of interpolation and the dots in the line correspond to the initial parameters for level $p \ge 3$. 	
	}\label{interpolation}
\end{figure}

\section{Our strategies}\label{sec:our_job}

In this section, we present MLI and greedy-MLI that can generate good initial parameters for QAOA and reduce most optimization rounds in an optimization run.

\subsection{MLI}\label{sec:MLI}

MLI makes attempts to generate the initial parameters at level $q_{0}+l$ by applying consecutive interpolations into the optimized parameters at level $q_{0}$, where $2 \le q_{0} \le p-2$ and $1 < l \le p-q_{0}$, and we refer to this operation as multilevel~leapfrogging~initialization. After initialization, MLI optimizes the initial parameters at level $q_{0}+l$, then update $q_{0}=q_{0}+l$. Repeatedly, until getting the optimized parameters at the target level $p$. By an operation of multilevel leapfrogging initialization, MLI can omit all the optimization rounds from level $q_{0}+1$ to $(q_{0}+l-1)$. Therefore, MLI executes optimization at few levels of the PQC rather than each level in an optimization run, and we call it multilevel~leapfrogging~optimization. For simplicity, we introduce a list $q$ to store the levels at which MLI performs optimization, where the elements of $q$ are incremental. The corresponding routine of MLI is in \textbf{Algorithm~\ref{MLI}}, where $j$ starts from 0. In addition, we give \textbf{Figure~\ref{interpolation}} to help readers understand the process of generating initial points by taking advantage of consecutive interpolations.

\medskip
The elements in the list $q$ are relevant to the depth step $l$. For example, the list $q = [1,\cdots,q_{0},p]$ when $l = p-q_{0}$ and $q=[1,2,\cdots,p]$ when $l$ is the constant one (i.e., INTERP). The 
number of optimization rounds in an optimization run decreases as $l$ increases. In other words, MLI may reduce more costs for the outer loop optimization as $l$ increases. However, there is a trade-off between the performance and the costs of MLI. MLI may have lower performance when $l$ is greater than a certain value if there is not enough prior information to produce high-quality initial parameters for the larger circuit. So far, the performance of MLI and the effects of $l$ on the performance of MLI are unknowable, which are questions to be investigated in the following work.
\begin{figure*}[ht]
	\centering
	\setlength{\abovecaptionskip}{0.cm}
	\subfigure[The comparison of OAR]{
		\includegraphics[width=0.45\textwidth]{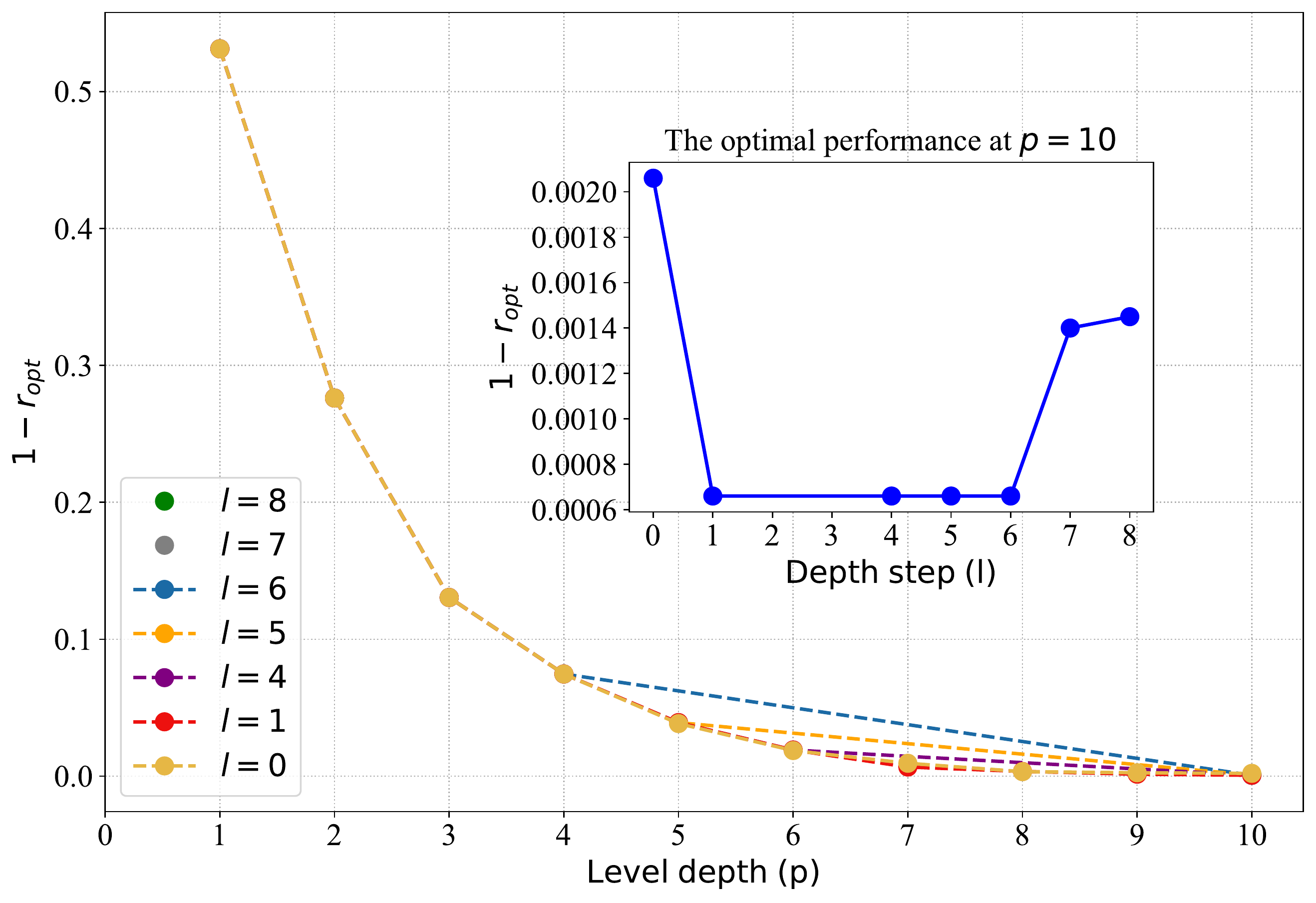}
		\label{l_ratio:OAR}
	}
	\subfigure[The comparison of AAR]{
		\includegraphics[width=0.45\textwidth]{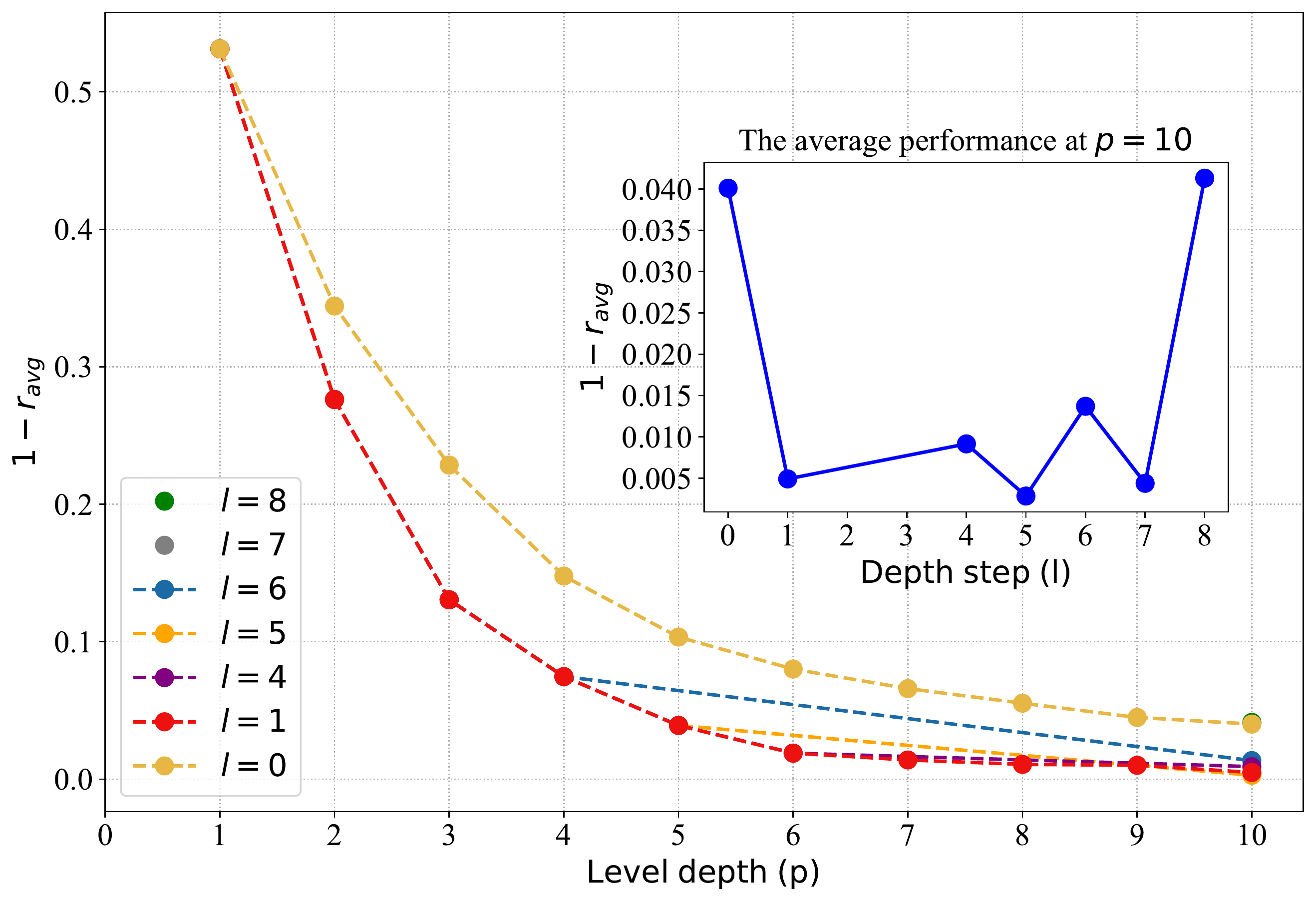}
		\label{l_ratio:AAR}
	}
	\caption{The AR of MLI versus the level depth and the depth step $l$, where $l = 0$ and $l = 1$ respectively correspond to RI and INTERP. For each $l$, MLI achieves better performance as $p$ increases. (a) The OAR of MLI at each level depth. At shallow level depth, the quasi-optimal solution obtained by 500 runs of RI is the same as that obtained by INTERP (the yellow and red dot coincide completely), but with the increase of $p$, the quasi-optimal solution obtained by RI may be slightly lower than that obtained by INTERP. The inset depicts the OAR of MLI versus the depth step when $p=10$. We note that MLI with $l = 4,5$ can obtain the identically optimal performance as INTERP at the target level, and the OAR obtained by RI is slightly lower than the others when $p = 10$. (b) The AAR of MLI versus the level depth, where the yellow and green dot are partly coincident when $p = 10$.  The inset reflects the effects of $l$ on the AAR obtained by MLI at $p = 10$, and it shows that MLI with $l = 5$ can get a higher AAR than INTERP. However, the average performance of MLI with $l = 8$ is worse than INTERP and RI, suggesting too large $l$ can lead to a decline in the performance of MLI.}
	\label{l_ratio}
\end{figure*}

\begin{algorithm}[H]
	\renewcommand{\algorithmicrequire}{\textbf{Input:}}%¸ü¸ÄÊäÈëÃû³Æ
	\renewcommand{\algorithmicensure}{\textbf{Output:}}%¸ü¸ÄÊä³öÃû³Æ
	\caption{ Multilevel Leapfrogging Interpolation strategy}\label{MLI}
	\begin{algorithmic}[1]
		
		\Require The target level $p$, the list $q$ that contains the levels where MLI executes optimization during an optimization run, random initial parameters $(\boldsymbol{\gamma_{i}}^{0}  ,\boldsymbol{\beta_{i}}^{0})$ at level $i = q[j]$, where $j$ starts from 0 and $q[j] < q[j+1]$.
		
		\Ensure The optimized parameters $(\boldsymbol{\gamma_{p}}^{*}  ,\boldsymbol{\beta_{p}}^{*})$ at level $p$.
		
		\State Optimize initial parameters $(\boldsymbol{\gamma_{i}}^{0}  ,\boldsymbol{\beta_{i}}^{0})$ to get optimized parameters $(\boldsymbol{\gamma_{i}}^{*}  ,\boldsymbol{\beta_{i}}^{*})$.
		
		\State Update the index $ j = j+1 $ and generate the guesses of initial parameters for level $q[j]$ by applying linear interpolation to the optimized parameters at level $q[j-1]$.
		
		\State  Repeat $1-2$ until getting optimized parameters of the target level.
	\end{algorithmic}
\end{algorithm}

In our preliminary work, we lack prior experience in determining the value of $l$. To circumvent the need for optimization at higher levels, we explicitly set $l = p-q_{0}$ and derive initial parameter estimates for the target level $p$ by applying consecutive interpolations to the optimized parameters at level $q_{0}$. In the following work, we take different values of $l$ to investigate the effects of $l$ on the performance of MLI. We aim to find $l$ that makes MLI reduce optimization rounds without sacrificing too much performance. What needs to be mentioned is that the circuit with $q_{0}$-level ansatz is still trained by INTERP, and we refer to it as the pre-trained circuit. Thus, the list $q = [1,2,\cdots,q_{0},p]$ in this case. For the following simulations, we first randomly generate several 3-regular graphs with vertex numbers $n=10, 12$ by the Python library networkx. Then, we explore the performance of MLI on given graphs when $l$ takes different values. In our simulations, we set $p=n$ and respectively perform 500 optimization runs INTERP and MLI, where they start from the same initial parameters at the first level. In addition, we also perform 500 optimization runs of RI at each $p$ to compare the performance of RI and INTERP (MLI) under the same level depth.We set the maximum number of iterations for the optimization at different level depths and make them our stop conditions. In this section, we analyze the effects of $l$ on MLI performance by taking a 3-regular graph with $n = 10$ as an example. The additional results on a graph with $n = 12$ are given in Appendix~\ref{AR_l_12}. For ease of expression, the corresponding value of $l$ is represented as $l^{a}$ when MLI can get the same quasi-optimal solution as INTERP at the target level. The corresponding pre-trained layer of $l^{a}$ consists of $q_{0}^{a}$-level ansatz. 

\begin{figure}[t]
	\centering
	\setlength{\abovecaptionskip}{0.cm}
	\subfigure{
		\includegraphics[width=0.45\textwidth]{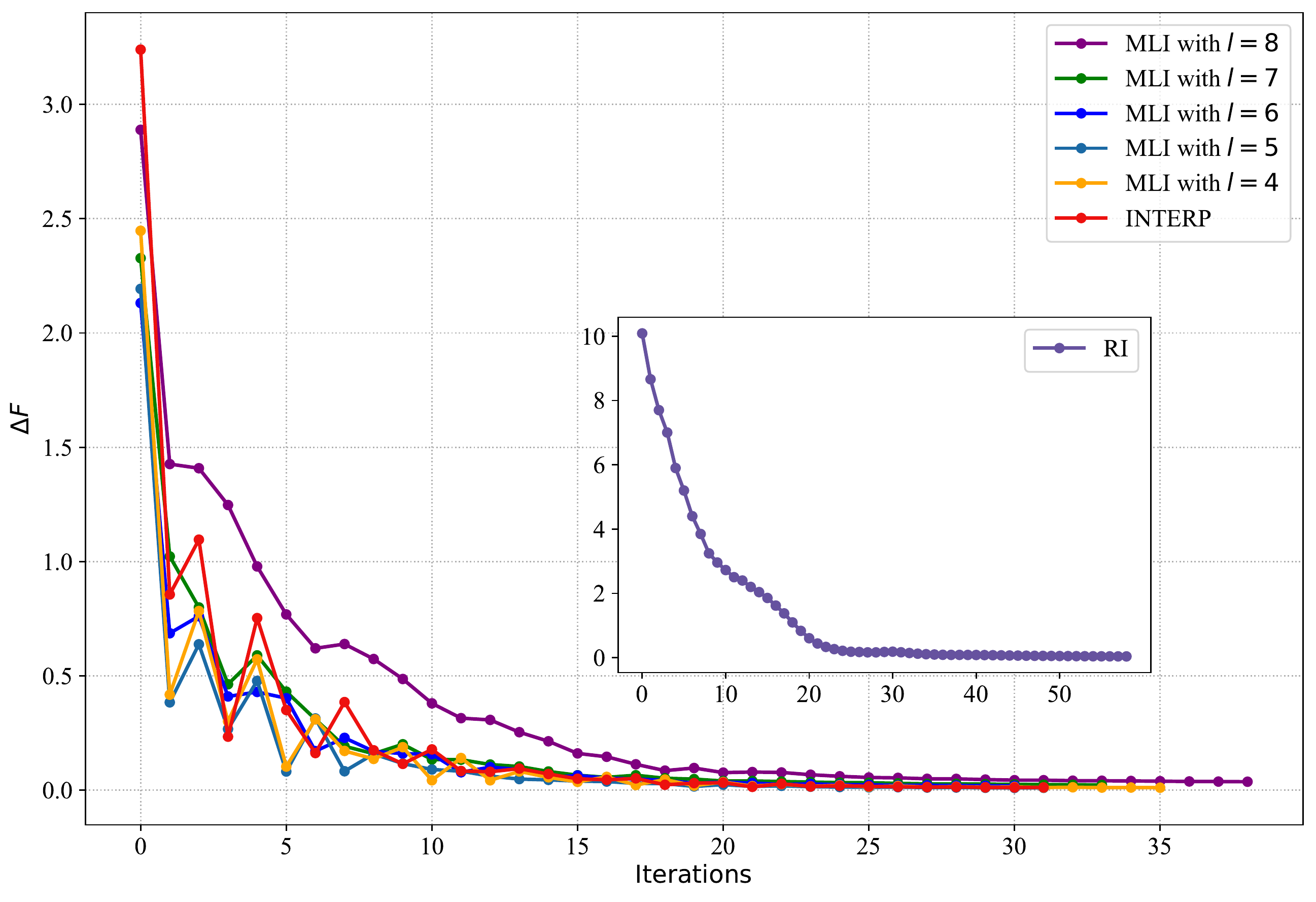}
	}
	\caption{The changes of $\Delta F$ versus the number of iterations when the initial parameters at $p = 10$ can converge to a quasi-optimal solution after optimization, where $\Delta F$ is the difference between the current expectation function value and the maximum expectation value $F_{max}$. The results demonstrate that the convergence speed of the function is similar even though the PQC starts from different initial parameters generated by multilevel leapfrogging initialization or level-by-level. In 500 optimization runs of RI when $p = 10$, there is one set of random parameters that can converge to its quasi-optimal solution, and the changes of $\Delta F$ versus the number of iterations are shown in the inset. The initial expectation value shows that random parameters are far from the quasi-optima. Thus, more iterations are required. In this figure, the consumed iterations are respectively 59 (RI), 39 (MLI with $l = 8$), 34 (MLI with $l = 7$), 31 (MLI with $l = 6$), 32 (MLI with $l = 5$), 36 (MLI with $l = 4$), and 32 (INTERP).}
	\label{l_convergence}
\end{figure}

\medskip
\textbf{Figure~\ref{l_ratio}} mainly plots the optimal AR (OAR) and average AR (AAR) obtained by MLI versus the level depth $p$ and the depth step $l$. From the simulation results, we observe that the AR gradually approaches one as $p$ increases. MLI consistently attains the nearly identical OAR as INTERP under $l^{a} = 4,5,6$ and gets a similar AAR with INTERP when $l = 5, 7$. These results affirm the initial success of MLI. That is, there are some values of $l(>1)$ that allow MLI to obtain the quasi-optimal solution and a similar average performance to INTERP by consuming fewer optimization rounds. Meanwhile, to study the convergence speed of the expectation function, we plot the detailed changes of $\Delta F$ versus the iterations when QAOA can get the quasi-optimal solutions at the target level depth. During optimization, we observe that the function value can reach convergence before the maximum number of iterations. To show the specific consumption of iterations by various strategies, we allow the optimization to end earlier if the expectation function value satisfies Equation~\ref{stop_condition}, where $\Delta = 0.001$. From these simulation results in \textbf{Figure~\ref{l_convergence}}, we find the expectation function gradually converges towards $F_{max}$ as the number of iterations increases, and QAOA consumes fewer iterations to reach convergence at $ p = 10$ when it starts with parameters generated by interpolation instead of random parameters. More surprisingly, we find that the convergence speed of the function remains relatively consistent even though QAOA starts with different initial parameters generated by MLI or INTERP. In other words, some initial parameters generated by multilevel leapfrogging initialization can converge to the quasi-optimal parameters at $ p = 10$, and the consumed iterations are almost the same as INTERP. Besides, it is essential to highlight that the bigger values of $l$ are not always better since the large values of $l$ may lead to a decline in the performance of MLI, such as slower convergence or potential entrapment in local optima, as evident in both Figure~\ref{l_ratio} and ~\ref{l_convergence}. Overall, these findings suggest that MLI can get the quasi-optimal solution on par with INTERP while operating at a lower expenditure.

\medskip

\begin{figure}[htbp]
		\centering
		\setlength{\abovecaptionskip}{0.cm}
		\subfigure{
			\includegraphics[width=0.45\textwidth]{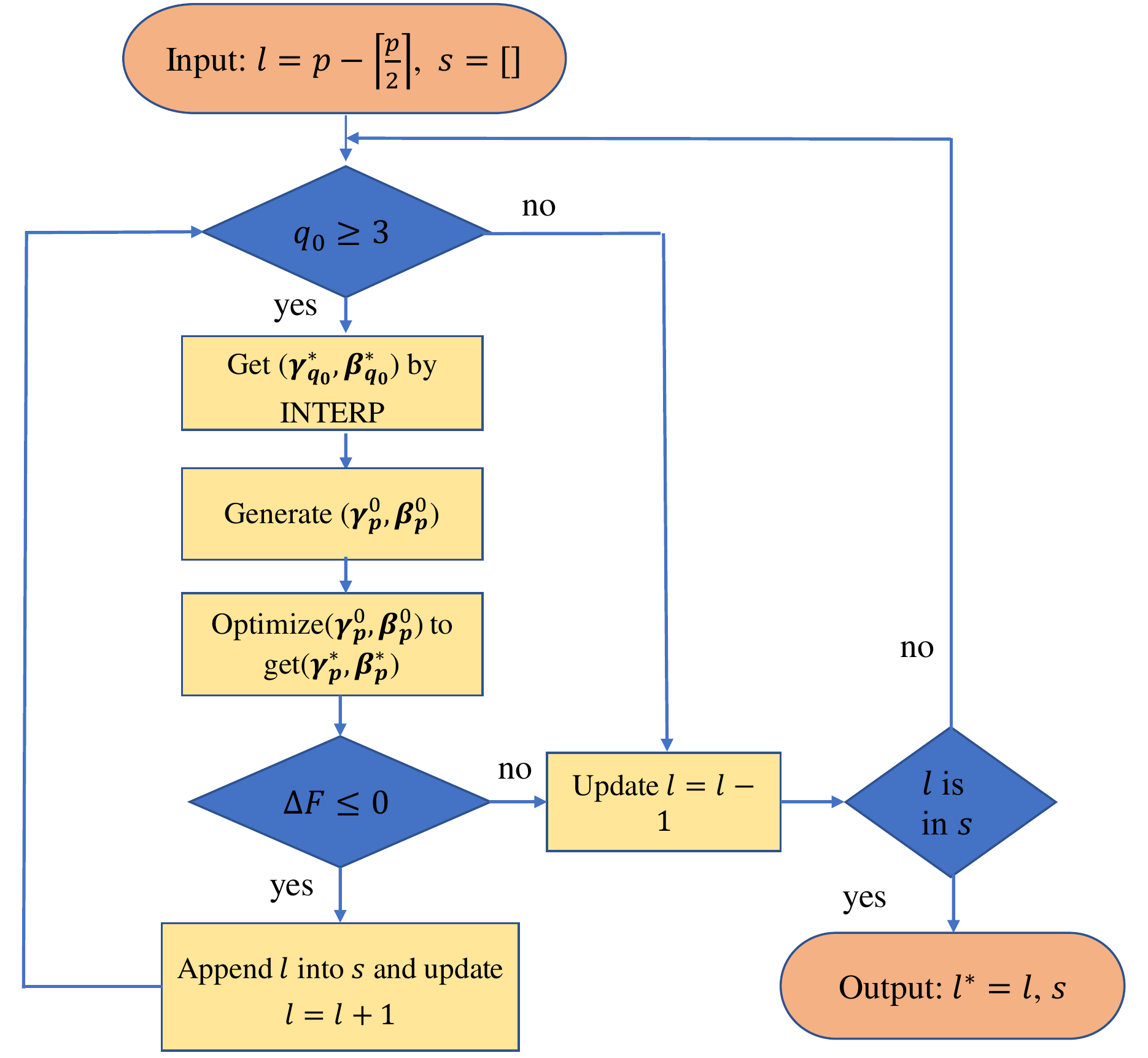}
		}
		\caption{The procedure of searching for $l^{*}$, where we set $\Delta F$ equals the difference between $F(\boldsymbol{\gamma_{p}}^{m},\boldsymbol{\beta_{p}^}{m})$ and the quasi-optimal solution (instead of $F_{max}$). The list $s$ is empty at the beginning. During the process of searching for $l^{*}$, the list $s$ stores the values of $l^{a}$. MLI with $l^{a}$ can achieve the same quasi-optima as INTERP by O[poly($p$)] optimization runs. The $l^{*} $ is the largest value in the list $s$. }\label{search_l}	
	\end{figure}

There are different values of $l^{a}$ in the above instances,  and we denote $l^{a}$ as $l^{*}$ when MLI with $l^{a}$ can save the most optimization rounds and get an identical optimal performance to INTERP. We can find $l^{*}$ by the following steps: (i) Assume the initial value of $l$ is $p -\left \lceil \frac{p}{2} \right \rceil$. (ii) Get the optimized parameters at level $q_{0}$ by INTERP, where $q_{0}=p-l$. (iii) Produce the initial parameters $(\boldsymbol{\gamma_{p}}^{0},\boldsymbol{\beta_{p}^}{0})$ by applying successive linear interpolations to $(\boldsymbol{\gamma_{q_{0}}^}{*},\boldsymbol{\beta_{q_{0}}^}{*})$, then optimize $(\boldsymbol{\gamma_{p}}^{0},\boldsymbol{\beta_{p}^}{0})$ to get the expectation value at the target level. (iv) Calculate the difference $\Delta F$ between $F(\boldsymbol{\gamma_{p}^{m}},\boldsymbol{\beta_{p}^{m}})$ and the quasi-optimal solution obtained by performing $c$ optimization runs of INTERP, where $F(\boldsymbol{\gamma_{p}}^{m},\boldsymbol{\beta_{p}^}{m})$ is the largest expectation value in $c$ optimization runs of MLI and $c~$ is the polynomial level of $p$. Append the value of $l$ into the list $s$ if $\Delta F \le 0 $, then increment $l$ by one. If $\Delta F >0 $ (i.e., the optimal performance of MLI with $l$ is worse than that of INTERP), update $l = l-1$. In this step, our primary objective is to search for better values of $l^{a}$ that further reduce the costs of INTERP in an optimization run without sacrificing its optimal performance. Thus, we set $\Delta F$ equals the difference between $F(\boldsymbol{\gamma_{p}}^{m},\boldsymbol{\beta_{p}^}{m})$ and the quasi-optimal solution of INTERP (instead of $F_{max}$), which is different from the definition in \text{ Figure~\ref{l_convergence}}. (v) Repeat (ii)–(iv) until $q_{0} \le 2$. (vi) Select the largest element from the list $s$. That is the value of $l^{*}$ that saves the most optimization rounds while guaranteeing the performance of MLI. The detailed process of finding $l^{*}$ is shown in \textbf{Figure~\ref{search_l}}.

\begin{figure}[htbp]
		\centering
		\setlength{\abovecaptionskip}{0.cm}
		\subfigure{
			\includegraphics[width=0.45\textwidth]{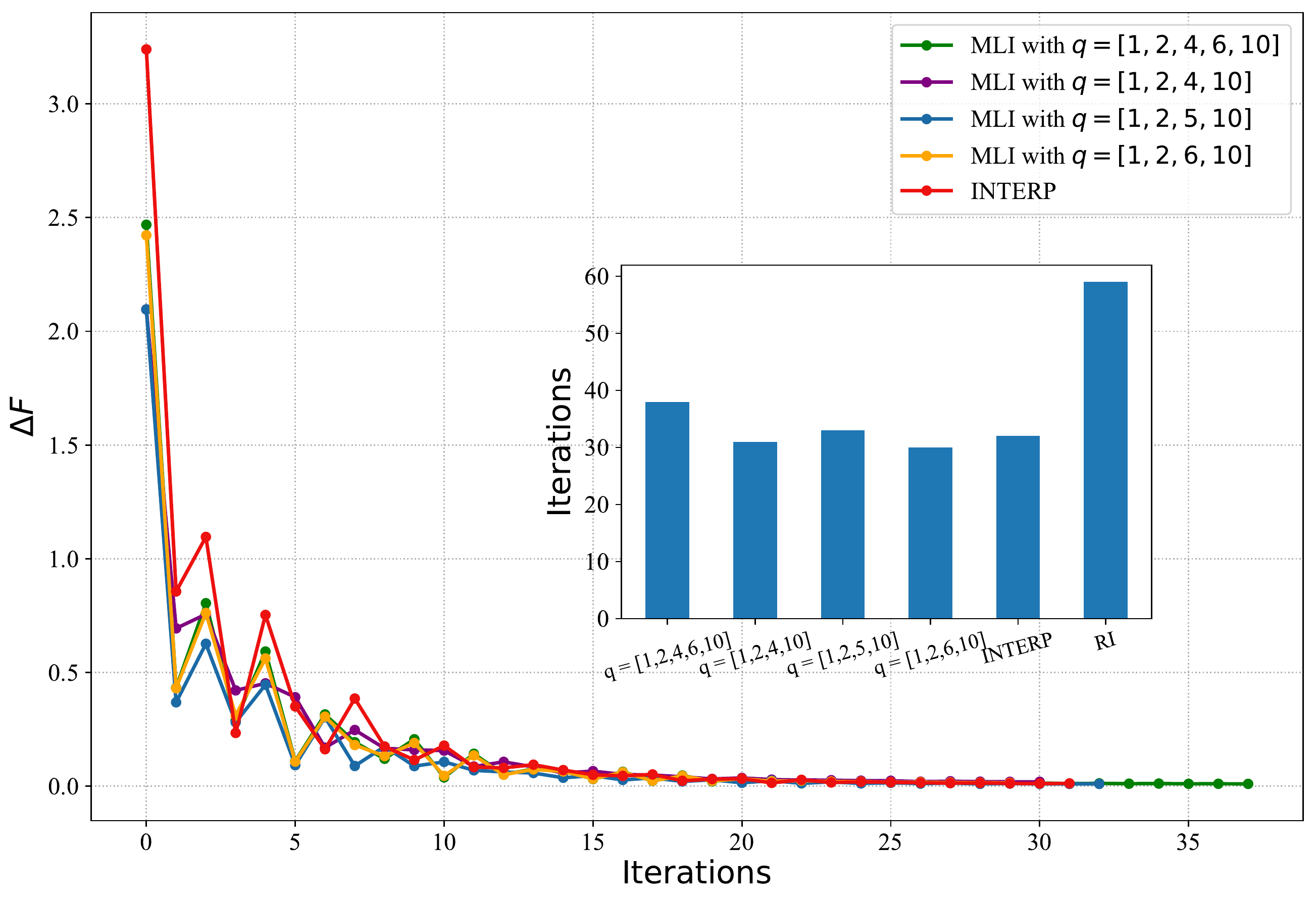}
		}\caption{The changes of $\Delta F$ versus the number of iterations when the initial parameters at $p = 10$ can converge to a quasi-optimal solution after optimization, where $\Delta F$ is the difference between the current expectation function value and the maximum expectation value $F_{max}$. The inset is to depict the detailed consumption of iterations. From these results, it is clear that the expectation function can converge to the same quasi-optimal solution after optimization at a similar convergence speed even though QAOA starts from different initial parameters that are produced by various simplifications of $q$.}
		\label{q_convergence}
	\end{figure}

\medskip
For the above instance, MLI reduces $(l^{a}-1)$ rounds of optimization beyond getting the quasi-optimal solution. However, getting the optimized parameters at level $q_{0}^{a}$ by INTERP still requires numerous costs when $q_{0}^{a}$ is large. To further cut down on costs during the optimization of MLI, we adopt a recursive step to simplify $q = [1, 2,\cdots,q_{0}^{a},p]$: (i) Reset the target level to $p = q_{0}^{a}$. (ii) Use the above preliminary work about MLI to search for the new value of $q_{0}^{a}$. (iii) Repeat (i) and (ii) until $p\le 3$. In our subsequent work, we simplify $q = [1, 2,\cdots, q_{0}^{a}, p]$ by the above recursive idea and perform 500 optimization runs of MLI, then we utilize AR to evaluate the performance of MLI with the simplified $q$. The comparable results in \textbf{Table~\ref{optimal_MLI}} suggest that MLI can get the quasi-optima at $p=10$ by the simplified $q$, such as $q=[1, 2, 4, 10]$. At this time, MLI saves about half the number of rounds of optimization compared with INTERP. From the numerical results in \textbf{Table~\ref{avg_MLI}}, we find there are some lists of $q$ that allow MLI to achieve a similar average performance to INTERP beyond getting the same quasi-optima as INTERP by consuming fewer optimization rounds. In addition, we delve into studying the convergence speed of the expectation function when QAOA commences from various initial parameters at level $p=10$ generated by different simplifications of $q$. Identically, we allow the optimization to end earlier when the expectation function value satisfies Equation~\ref{stop_condition}, where $\Delta = 0.001$. More detailed results are respectively shown in \textbf{Figure~\ref{q_convergence}}. Evidently, some initial parameters derived from different simplifications of $q$ can converge to quasi-optimal parameters after optimization. More importantly, the convergence speed of the expectation function has little difference. To be more specific, skipping optimizations at some levels may not significantly compromise the quality of initial parameters at the target level. Thus, the initial parameters generated by multilevel leapfrogging initialization might also exhibit high quality, thus improving the efficiency of MLI.

\begin{table*}[htbp]
	\centering
	\caption  { \centering The comparison of OPR among RI, INTERP and MLI with various $q$}
	\label{optimal_MLI}%
	\renewcommand\arraystretch{1.15}
	\begin{tabular}{@{}ccccccc@{}}
		\hline
		\multirow{2}*{Level} &\multirow{2}*{RI}&\multirow{2}*{INTERP} &\multicolumn{4}{c}{MLI}   \\ \cline{4-7}
		& & &$q=[1,2,4,6,10]$  &$q = [1,2,4,10]$ & $q=[1,2,5,10]$ &$q=[1,2,6,10]$ \\ \hline
		$p=1$ &0.4688   & 0.4688    &0.4688  & 0.4688   &0.4688 &0.4688\\
		$p=2$ &0.7238   & 0.7238   &0.7238   & 0.7238    &0.7238 & 0.7238\\
		$p=3$ &0.8695   &0.8694     &               &                 &        &\\
		$p=4$ &0.9253   &0.9254   & 0.9254 &0.9254		  &        &  	  \\
		$p=5$ &0.9616   & 0.9609    &      &                  & 0.9609 &         \\
		$p=6$ &0.9813   & 0.9809   & 0.9809                    &        & 0.9809  \\
		$p=7$ &0.9906   & 0.9933   &                          &        &         \\
		$p=8$ &0.9967   & 0.9965   &  &                    &         \\
		$p=9$ &0.9975   & 0.9985   &                      &        &         \\
		$p=10$ &0.9979    & 0.9993 &0.9993   &0.9993 & 0.9993 &0.9993  \\		
		\hline
	\end{tabular}
	%\footnotetext{}
\end{table*}

\begin{table*}[htbp]
	\vspace{-1.0em}
	\centering
	\caption { \centering The comparison of AAR between INTERP and MLI with various $q$}
	\renewcommand\arraystretch{1.15}
	\label{avg_MLI}%
	
	\begin{tabular}{ccccccc}
		\hline
		\multirow{2}*{Level} &\multirow{2}*{RI} &\multirow{2}*{INTERP} &\multicolumn{4}{c}{MLI}   \\ \cline{4-7}
		&& &$q=[1,2,4,6,10]$ &$q=[1,2,4,10]$  & $q=[1,2,5,10]$ &$q=[1,2,6,10]$ \\ \hline
		$p=1$ &0.4688   & 0.4688   & 0.4688 & 0.4688   & 0.4688 &0.4688 \\
		$p=2$ &0.6559   & 0.7238  & 0.7238 & 0.7238   &0.7238 &0.7238\\
		$p=3$ &0.7716   & 0.8695   &                &                 &   &    \\
		$p=4$ &0.8523   & 0.9254  & 0.9085 & 0.9085 &		            &	  \\
		$p=5$ &0.8966   & 0.9609   &        &                  & 0.9434   &       \\
		$p=6$ &0.9199   & 0.9811   &0.9480  &                 &       & 0.9683  \\
		$p=7$ &0.9341   & 0.9860   &         &                  &        &        \\
		$p=8$ &0.9447   & 0.9892   &  &         &                      & \\
		$p=9$ &0.9551   & 0.9909   &         &                  &        &         \\
		$p=10$&0.9599    & 0.9951   & 0.9951 &0.9962  & 0.9916 &0.9859  \\
		\hline
	\end{tabular}
	%\footnotetext{}
\end{table*}

\medskip
MLI can get the quasi-optimal solution with various values of $q$ in the above instance, and MLI with $q^{*}$ cuts down on most optimization rounds while getting the identical quasi-optima to INTERP. The list $q^{*}$ consists of $q_{0}^{*}$ and the level $i = 1, 2, p$. We can get $q^{*}$ by the following recursive steps: (i) For the target level $p$, search for $l^{*}$ by Figure~\ref{search_l}, then get $q_{0}^{*} = p-l^{*}$. (ii) Reset the target level to $p = q_{0}^{*}$. (iii) Repeat (i) and (ii) until the target level $p \le 3$. In our work, we randomly generate 150 3-regular graphs with $n=6,8,10,12,14$ and seek the list $q^{*}$ based on them. From the numerical results, we find that $q_{0}^{*}$ approximates $\frac{p}{2} -1$ when the target level $p>4$ and $p$ is even, $q_{0}^{*}$ approaches $\frac{p-1}{2}$ when the target level $p>4$ and $p$ is odd, and $q_{0}^{*}$ equals 2 when $p = 4$. Though numerous costs are required to search for the simplified $q$ that are `` universally good '', the obtained regularities can provide an initial conjecture of the simplified $q$ for MLI. The simplified $q$ allows QAOA to bypass about half of the optimization rounds and decreases execution times, potentially making the external loop optimization more simple and efficient. Thus, exploring the conjecture about the simplified $q$ is worthwhile in the long run.

%The procedure of searching for $l^{*}$, where we set $\Delta F$ equals the difference between $F(\boldsymbol{\gamma_{p}}^{m},\boldsymbol{\beta_{p}^}{m})$ and the quasi-optimal solution (instead of $F_{max}$). \textcolor{blue}{The lists $s_{0}$ and $s$ are empty at the beginning. During the process of searching for $l^{*}$, the lists $s_{0}$ and $s$ store the values of $l^{a}$ and $l^{b}$, respectively. MLI with $l^{a}$ can achieve the same quasi-optima as INTERP at the target level depth. MLI with $l^{b}$ can get not only identical optimal performance but also similar average performance to INTERP. The $l^{*} = $ max$(s)$ represents the largest value in the list $s$. In this figure, $F_{avg, M_{l}}$ and $F_{avg, I}$ respectively represent the average expectation function value obtained by multiple runs of MLI with $l$ and INTERP
	
	\subsection{greedy-MLI}\label{sec:greedy-MLI}
	
	The smoothness of parameters improves the success of linear interpolation, increasing the probability of finding quasi-optimal solutions for INTERP and MLI. However, in practice, it is inevitable to encounter situations where optimized parameters at level $p$ oscillate back and forth, as observed in Ref.~\cite{bilinear}, and the performance of INTERP and MLI may slightly or dramatically deteriorate at the subsequent levels when they start from the initial points produced by non-smoothly optimized parameters. To tackle this problem, the greedy-MLI strategy is presented, and it tries to direct the optimization process towards better solutions by pruning less favorable parameter configurations. The Algorithm~\ref{greedy-MLI} outlines the detailed procedures of the greedy-MLI strategy.

	\medskip
	Using the observations in a numerical study, we note that non-smooth parameters tend to exhibit poorer or the same performance as smooth parameters, and the parameters with a high AR (i.e., higher performance) are more likely to follow the desired pattern compared with those with a small $r$. Leveraging the above observations, the greedy-MLI strategy is designed to effectively search for smooth parameters and obtain quasi-optimal solutions. Greedy-MLI executes optimization in parallel, starting from $c$ pairs of initial parameters \cite{multistart}. This parallel execution allows for the exploration of different regions of the parameter space simultaneously, increasing the likelihood of finding promising solutions. After each round of optimization, greedy-MLI only retains the optimized parameters with the highest AR, and there may be multiple sets of retained parameters. This selective preservation focuses on parameter configurations that offer better performance. The preserved parameters from the previous level are utilized to produce an educated guess for the next level of optimization. This approach leverages the knowledge gained from successful optimizations to guide the optimization process at higher levels, enhancing efficiency and reducing the chances of getting trapped in non-smooth parameter regions. It is worthwhile to note that all subsequent steps based on them (e.g., producing initial points for the next level and executing optimization) stop when the low-quality parameters are eliminated. This early termination prevents wasteful exploration of non-promising regions of the parameter space, conserving computational resources. Thus, greedy-MLI consumes fewer resources to get quasi-optimal parameters compared with $c$ optimization runs of MLI.
	
	\medskip
	Overall, the success of the greedy-MLI strategy lies in its ability to make informed decisions based on observed performance patterns, selectively preserving promising parameters, and leveraging parallel optimization to explore larger parameter space. More importantly, greedy-MLI terminates to explore non-promising regions of the parameter space in time, preventing wasting costs. These characteristics might make greedy-MLI a valuable tool for researchers and practitioners to solve other QAOA optimization tasks, offering improved performance and resource efficiency. 
	
	\begin{algorithm}[H]
		\renewcommand{\algorithmicrequire}{\textbf{Input:}}%¸ü¸ÄÊäÈëÃû³Æ
		\renewcommand{\algorithmicensure}{\textbf{Output:}}%¸ü¸ÄÊä³öÃû³Æ
		\caption{ Greedy multilevel leapfrogging interpolation strategy }\label{greedy-MLI}
		\begin{algorithmic}[1]
			
			\Require The target level $p$, the list $q$ that stores the levels where greedy-MLI executes optimization in an optimization run, $c$ pairs of random initial parameters $(\boldsymbol{\gamma_{i}}^{0}  ,\boldsymbol{\beta_{i}}^{0})$ at level $i= q[j]$, where $j$ starts from 0 and $q[j]<q[j+1]$.
			
			\Ensure The optimized parameters $(\boldsymbol{\gamma_{p}}^{*}  ,\boldsymbol{\beta_{p}}^{*})$ at level $p$.
			
			\State Optimize initial parameters $(\boldsymbol{\gamma_{i}}^{0}  ,\boldsymbol{\beta_{i}}^{0})$ in parallel and get $c$ pairs of optimized parameters $(\boldsymbol{\gamma_{i}}^{*}  ,\boldsymbol{\beta_{i}}^{*})$.
			
			\State Perform one round of selection to preserve $k(\ge 1)$ pairs of optimized parameters with the highest approximation ratio.
			
			\State Update $c = k$, $ j = j+1 $ and generate $c$ pairs of the initial parameters for level $i=q[j]$ by the interpolation technique.
			
			\State  Repeat $1-3$ until getting optimized parameters of the target level $p=q[j]$.
			
		\end{algorithmic}
	\end{algorithm}
	
	\section{Comparison among strategies }\label{sec:VS}

	\begin{figure*}[htbp]
		\centering
		\setlength{\abovecaptionskip}{0.cm}
		\subfigure[The comparison of OAR when $n = 8$]{
			\includegraphics[width=0.38\textwidth]{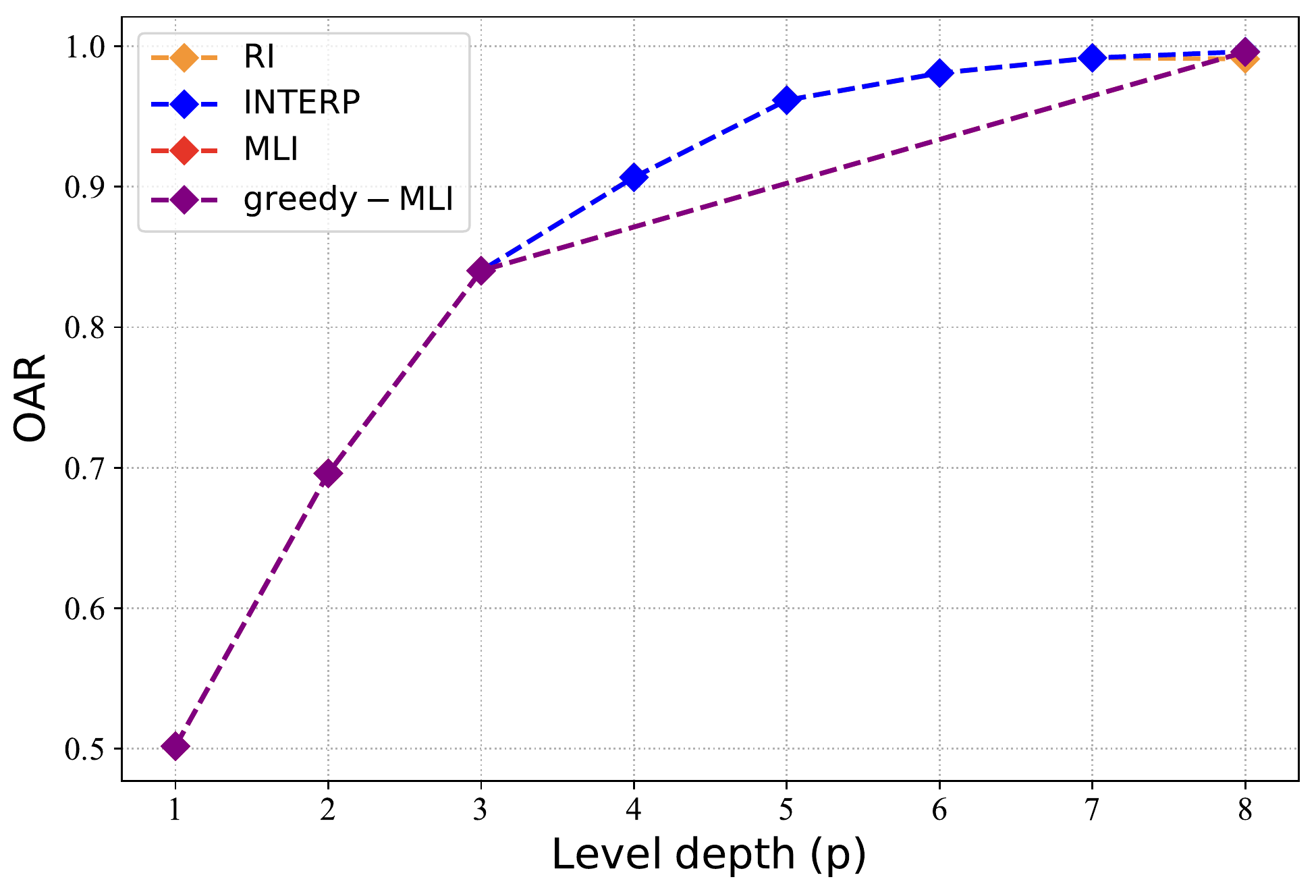}
		}
		\subfigure[The comparison of AAR when $n = 8$]{
			\includegraphics[width=0.38\textwidth]{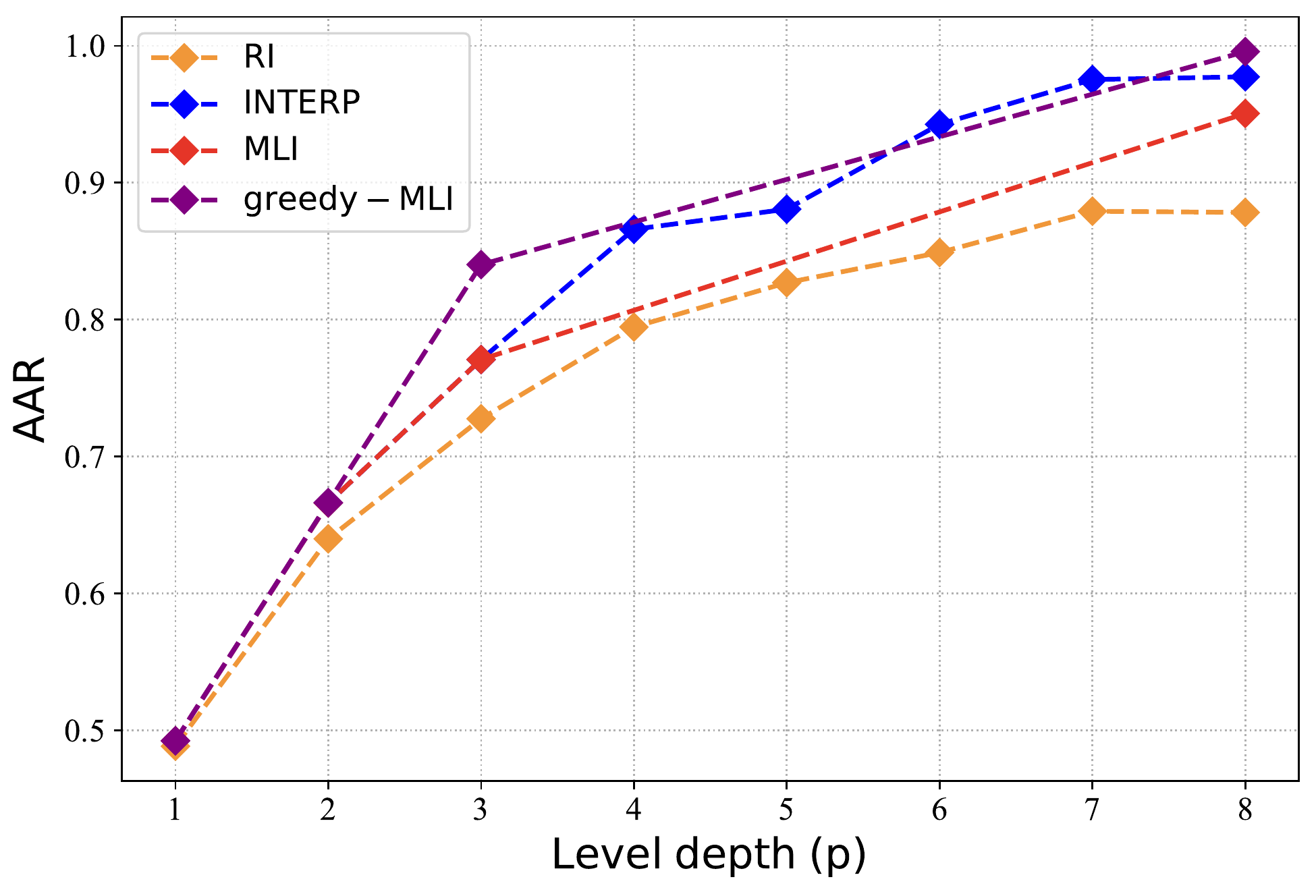}
		}
		
		\subfigure[The comparison of OAR when $n = 10$]{
			\includegraphics[width=0.38\textwidth]{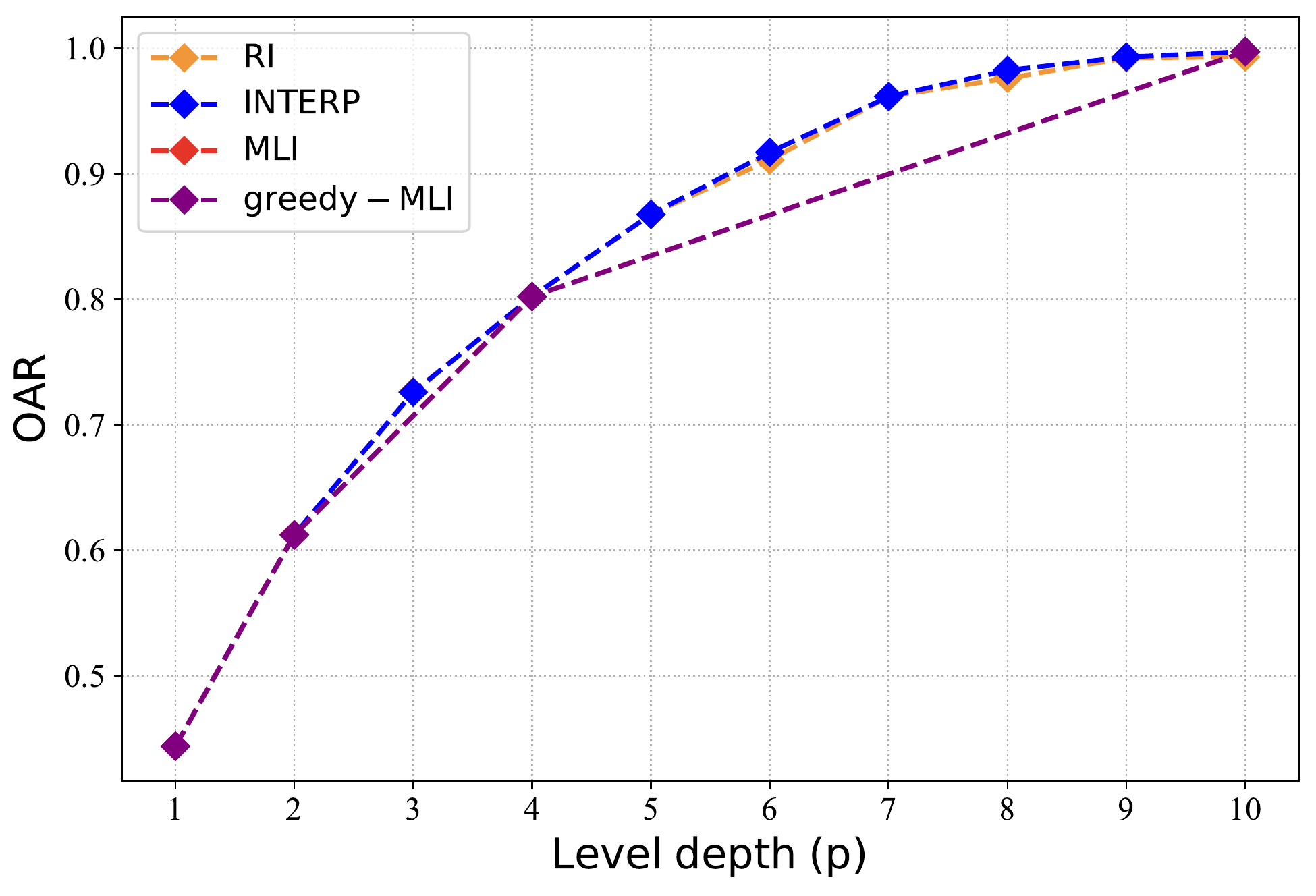}
		}
		\subfigure[The comparison of AAR when $n = 10$]{
			\includegraphics[width=0.38\textwidth]{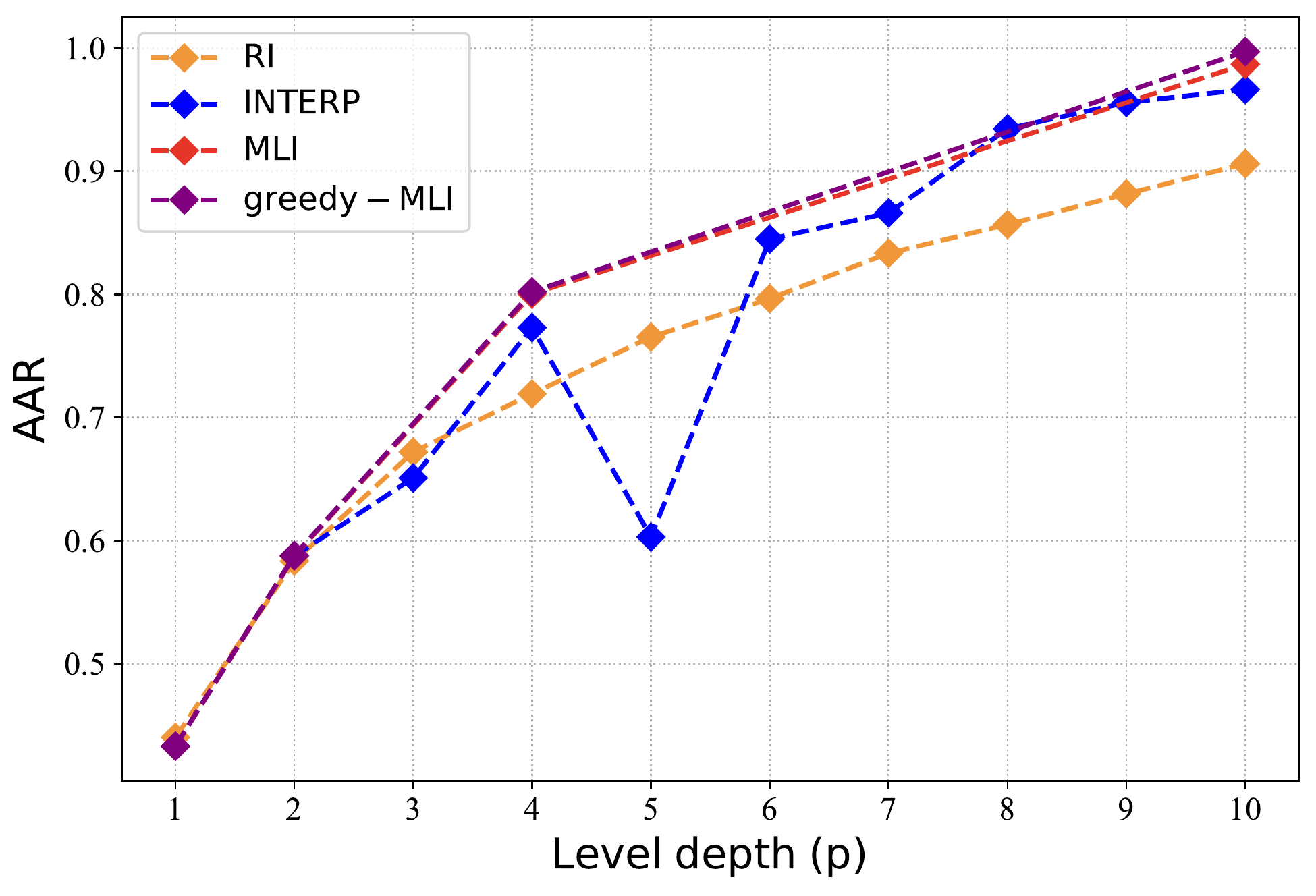}
		}
		
		\subfigure[The comparison of OAR when $n = 12$]{
			\includegraphics[width=0.38\textwidth]{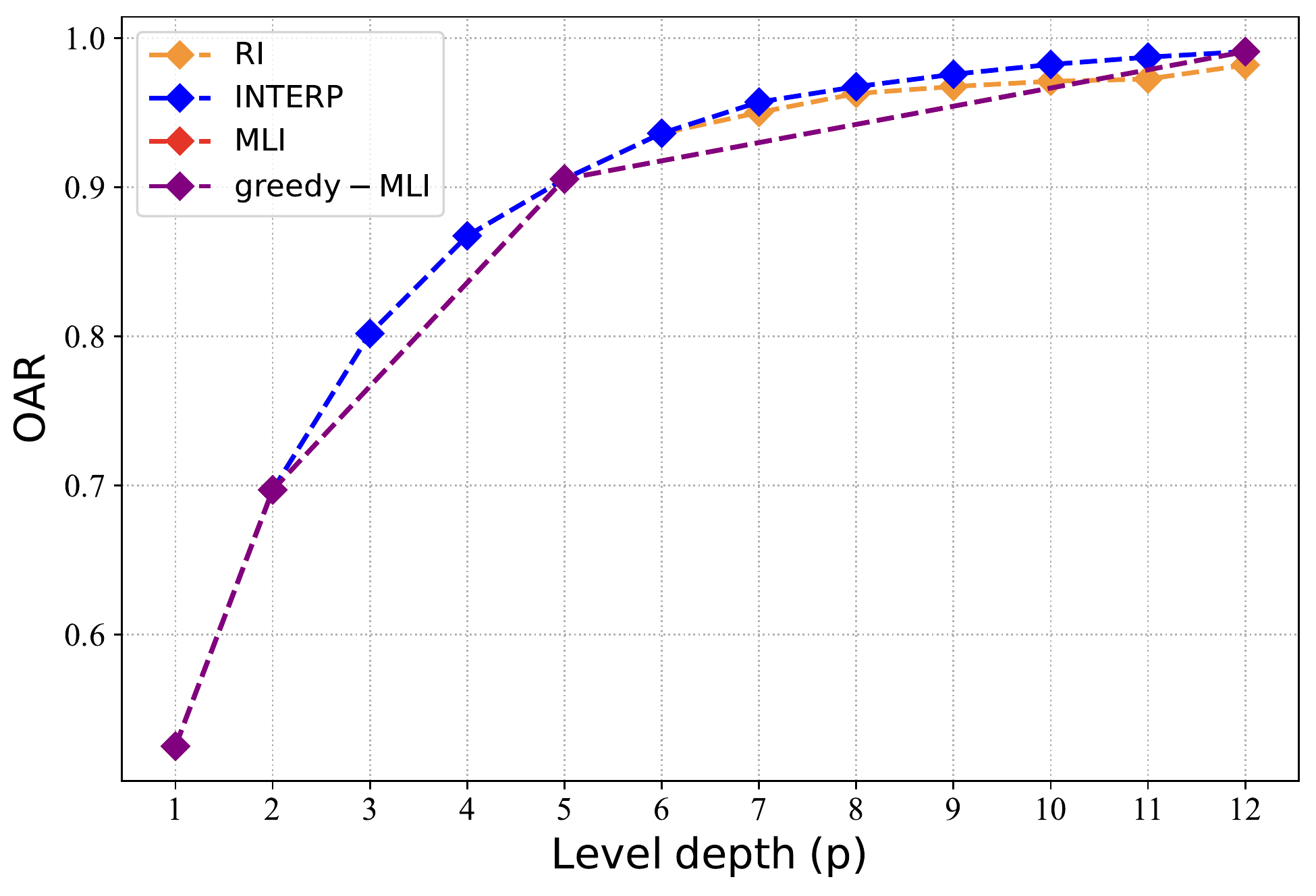}
		}
		\subfigure[The comparison of AAR when $n = 12$]{
			\includegraphics[width=0.38\textwidth]{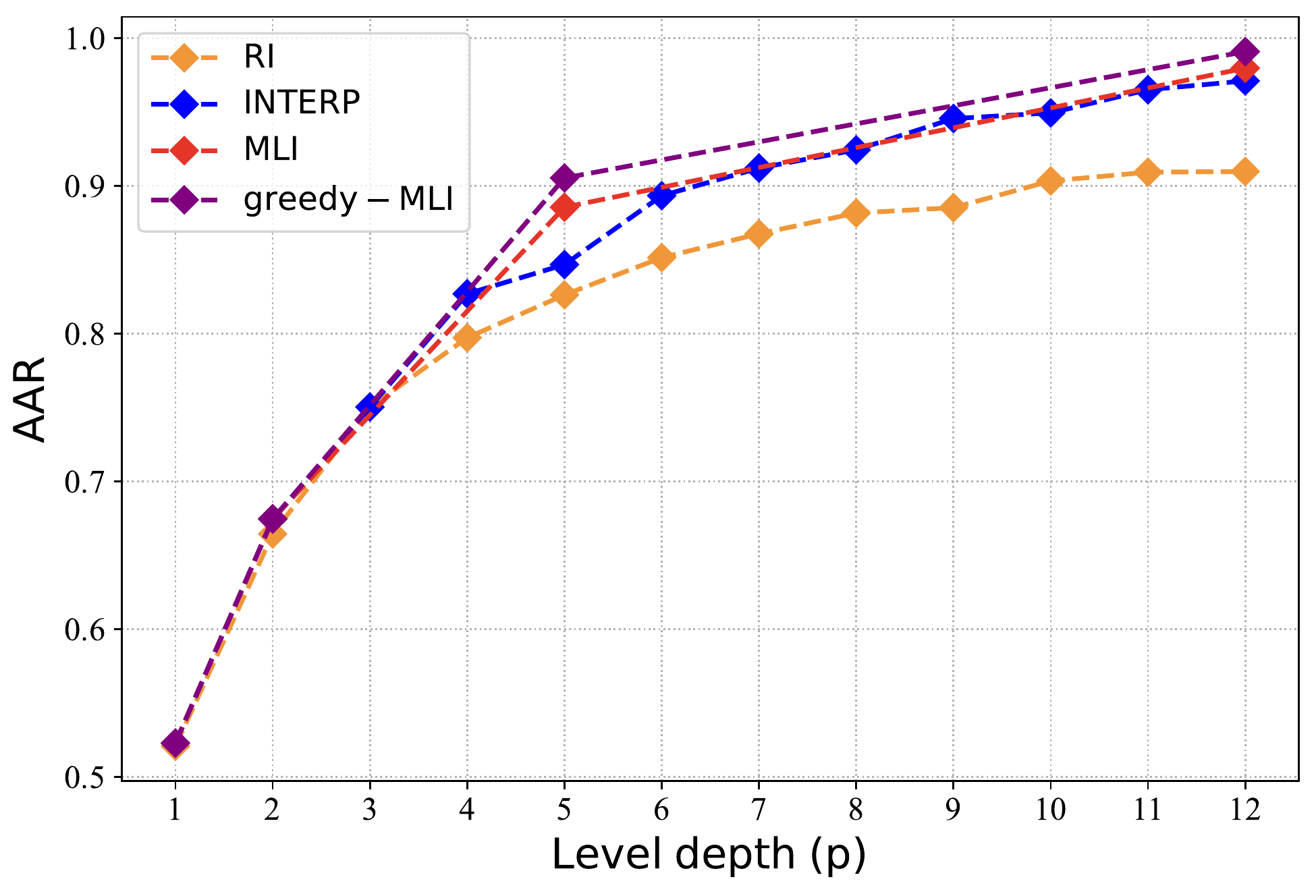}
		}
		\caption{The performance comparison of four strategies on 3-regular graphs with $n=8, 10, 12$. Here, we take (b), (c), and (d) as examples to analyze the simulation results. (b) The AAR of greedy-MLI is better than INTERP and MLI at level $p = 3$ ($p = 8$), which benefits from discarding a part of low-quality optimized parameters for level $p=2$ ($p=3$). (c) The OAR grows as level depth increases, and MLI and greedy-MLI can obtain the same quasi-optimal value as INTERP in fewer optimization rounds. In addition, the OAR obtained by 500 optimization runs of RI is the same as INTERP at shallow level depth. However, the OAR obtained by RI is gradually lower than that of INTERP as $p$ increases. (d) There is a sudden deterioration in the AAR of INTERP at $p = 5$. Upon analyzing the relevant data, we find that there are some parameters produced by applying interpolation to the non-smooth parameters, and QAOA may get stuck in low-quality solutions when it starts with these parameters, thus reducing the AAR.} 
		\label{AR_vs}
	\end{figure*}
	
	\begin{figure}[htb]
		\centering
		\setlength{\abovecaptionskip}{0.cm}
		\subfigure{
			\includegraphics[width=0.45\textwidth]{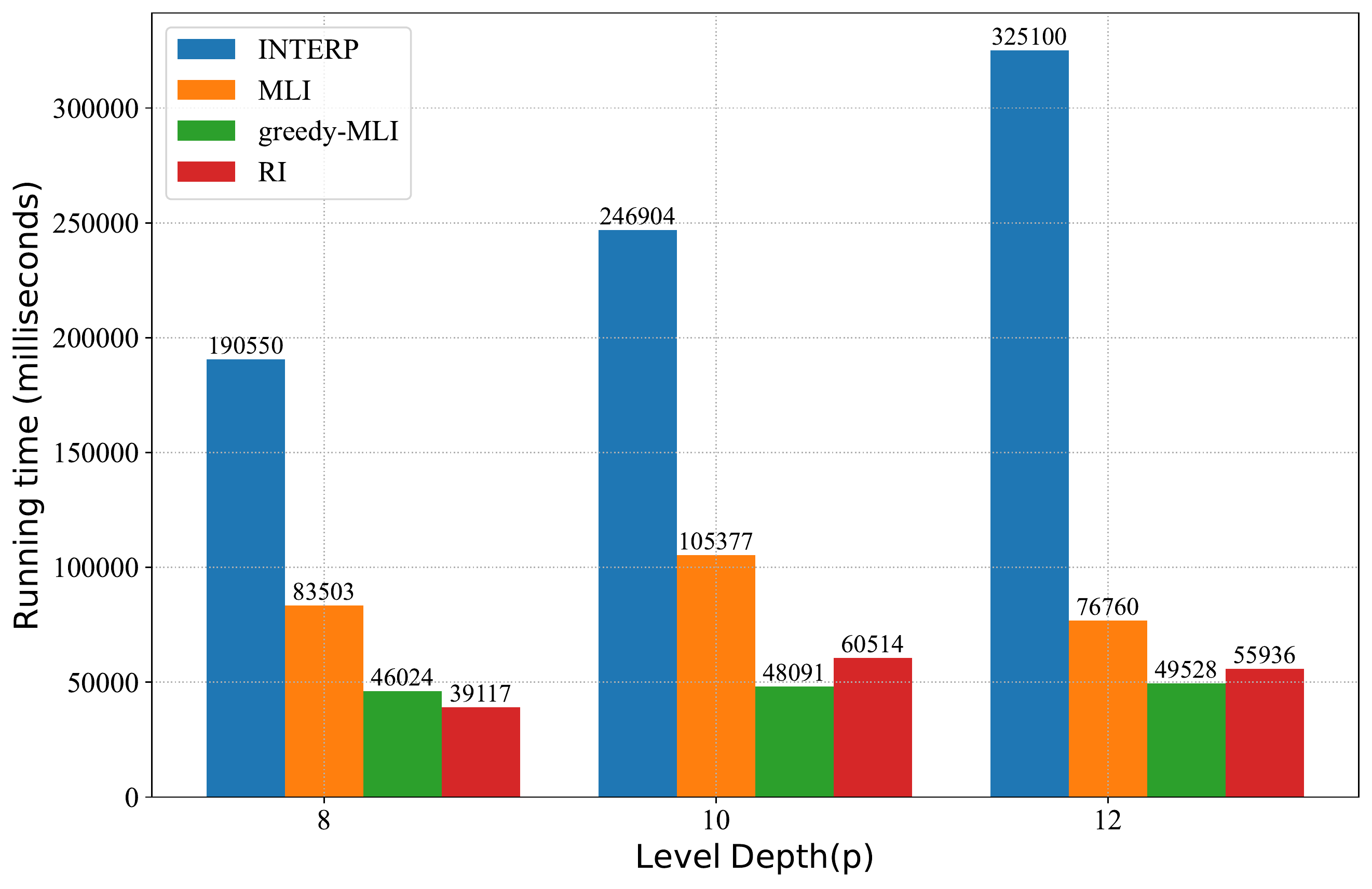}
		}
		\caption{The comparison of the total number of running time in 500 optimization runs of INTERP, MLI, RI, and an optimization run of greedy-MLI with $c = 500$. These results reveal a dramatic increase in the running time of INTERP and RI as the level depth grows. However, both MLI and greedy-MLI exhibit greater efficiency by bypassing optimization at deeper levels compared with INTERP, resulting in notable time savings. From the time consumption of INTERP, it is evident that our strategies are more efficient. In essence, these strategies intelligently skip time-consuming computations at higher levels, leading to more effective optimization.}\label{estimate_time}
	\end{figure}
	
	We generate 30 3-regular graphs for every vertex number $n$ to examine the difference in performance among heuristic strategies on the given graphs, where $n = 8, 10$, and $12$. On each graph instance, we perform 500 optimization runs for INTERP and MLI and an optimization run for greedy-MLI with $c=500$, where the target level equals the vertex number. In our simulation experiments, we both execute 500 optimization runs of RI at each $p$ to compare the performance with the other strategies at the same level depth. Besides, we set $\Delta_{1} \le \Delta$ and $\Delta_{2} \le \Delta$ as our stop conditions, where $\Delta = 0.001$. The computational experiments are conducted on a dedicated server equipped with an Intel(R) Xeon(R) Sliver 4316 CPU running at 2.3GHz and 251GB of RAM, providing ample computational resources, and we conduct the numerical simulations using the MindQuantum library version 0.7.0 and Python 3.7. The optimization of QAOA parameters is performed employing the Adam optimizer, a widely used optimization algorithm in machine learning. This choice is made to enhance the efficiency of parameter tuning and improve the convergence of the QAOA during the numerical experiments.
	
	\medskip
	We plot the AR obtained by different strategies on given graphs. From our examples in \textbf{Figure~\ref{AR_vs}}, both MLI and greedy-MLI strategies work well for all the instances we examine. MLI and greedy-MLI achieve identical optimal performance as INTERP but with significantly fewer rounds of optimization. Besides, the AAR obtained by MLI is comparable to INTERP, while the AAR of greedy-MLI surpasses that of INTERP. These results demonstrate that MLI and greedy-MLI are more efficient in reaching quasi-optimal solutions by skipping several optimization rounds without sacrificing too much performance. The efficiency of finding the quasi-optima might make them promising alternatives for optimization tasks, particularly when resources are limited. In addition, we carefully compare the average performance obtained by these strategies. The AAR obtained by RI is the lowest at the target level. Notably, we observe that greedy-MLI outperforms INTERP and MLI at the same level depth except for $p = 1$. This improvement is attributed to greedy-MLI discards a portion of low-quality optimized parameters during the optimization run. By doing so, greedy-MLI focuses on more promising parameter configurations, leading to better overall performance. The selective approach of greedy-MLI in parameter optimization may provide a clear advantage over the other strategies, particularly when dealing with higher-level optimizations.
	
	\medskip
	To show the training costs reduced by our strategies, we estimate the running time based on the calculation method provided in Ref.~\cite{multistart, runtime}. Specifically, the number of experimental repetitions in an optimization run is given by $T\times M$, where $M$ is the number of repeated measurements in each iteration, and $T$ is the number of consumed iterations for convergence. The time duration of a single repetition is considered to be $t$. Therefore, the time cost of performing an optimization round equals $t\times M \times T$. In \textbf{Appendix~\ref{total_run_time}}, we give the reason of $t$ equals one microsecond and the detailed methods of calculating the running time in an optimization run for each strategy. In our numerical simulation experiments, we set $M = 1000$, which means 1000 measurements of the PQC circuit are needed to calculate the expectation function value in each iteration. Then, we calculate the running time of 500 optimization runs of INTERP, MLI, RI, and an optimization run of greedy-MLI with $c = 500$, where the total running time of INTERP (RI and MLI) is the accumulation of the time consumption in each optimization run. More details are shown in \textbf{Figure~\ref{estimate_time}}.
	
	\medskip
	These results in Figure~\ref{estimate_time} suggest that INTERP and RI consume more running time than MLI and greedy-MLI, and the total running time of INTERP dramatically increases as the target level depth grows. The selection of initial parameters can affect $T$. For RI, QAOA tends to require more iterations to reach convergence when the random initial parameters are far away from the quasi-optimal solution, thus increasing the consumption of time. While for INTERP, its high consumption of running time mainly lies in executing optimization at each level depth $i$. Although the running time of RI is less than INTERP and our strategies when $p = 8$, the time consumed by greedy-MLI is gradually less than RI as $p$ increases. Compared with random parameters, the initial parameters generated by greedy-MLI may require fewer iterations to converge to the quasi-optimal parameters. Besides, greedy-MLI discards several non-promising parameters during optimization, thus decreasing the optimization rounds. These advantages may remedy the consumption of several optimization rounds in an optimization run, thus allowing the performance of greedy-MLI to surpass RI. In addition, we note the running time reduced by MLI and greedy-MLI for INTERP grows as $p$ increases, which benefits from reducing more rounds of optimization at the larger level by increasing the depth step $l$, and the advantages of MLI and greedy-MLI are notable at $p = 12$ compared with $p = 8,10$. The consumption of iterations has a significant impact on the overall running time. In our study, we have also generated a plot illustrating the total number of iterations across multiple optimization rounds in relation to the level depth. Further details can be found in \textbf{Appendix~\ref{total_ITR}}.
	
	\medskip
	Overall, the analysis of simulation results highlights the superiority of MLI and greedy-MLI over INTERP and RI in terms of efficiency and performance. MLI (greedy-MLI) can achieve identical OAR as INTERP but with significantly fewer rounds of optimization and 1/2 (1/5) of running time. Especially for greedy-MLI, it not only achieves comparable quasi-optimal values with fewer optimization rounds but also demonstrates a higher average performance by discarding less promising parameter configurations. These findings underscore the effectiveness of M-Leap and the selective approach of greedy-MLI.

	\section{Conclusion and outlook}\label{conclusion}
	
	In this paper, we present two heuristic strategies to generate initial QAOA parameters for the PQC. The central idea behind them is to generate initial parameters for level $p+l$ ($l>1$) leveraging the optimized parameters obtained at level $p$, a concept we refer to as multilevel leapfrogging initialization. The final result is that MLI and greedy-MLI execute optimization at few levels rather than each level, significantly reducing the number of optimization rounds required in the external loop optimization. Though this paper mainly considers QAOA, the idea of M-Leap may suit a broader class of problems,  such as quantum circuit design, quantum circuit learning, quantum neural networks, and so on. It is interesting to explore the realistic performance of M-Leap on these training tasks.
	\medskip
	
	In our work, we explore the performance of our strategies on the Maxcut problem. In the study of MLI, we focus on investigating the effects of $l$ on the performance of MLI. From the numerical results, we find that the performance of MLI may lower when $l$ surpasses $l^{*}$ if there is not enough pre-information to generate good initial points. In addition, we conclude the regularities of the value of $l^{*}$ through numerous benchmarking. The obtained regularities can provide an initial conjecture of $q$ that allows MLI to save most optimization rounds, potentially making the classical outer learning loop more simple and efficient. The simulation results show that MLI can get the quasi-optimal solution while consuming only 1/2 of the running costs required by INTERP. The success of MLI and INTERP is related to the smooth parameter pattern. During the simulation, we note that non-smooth parameters $(\boldsymbol{\gamma_{p}}^{n}  ,\boldsymbol{\beta_{p}}^{n})$ may show the same optimal performance as smooth parameters $(\boldsymbol{\gamma_{p}}^{s}  ,\boldsymbol{\beta_{p}}^{s})$. However, the performance of INTERP and MLI may deteriorate at the subsequent levels when they start from the initial points produced by non-smoothly optimized parameters. To overcome this shortcoming, we present greedy-MLI, which is an extension of MLI. To search for the smooth parameters during the optimization, greedy-MLI executes parallel optimization starting from $c$ pairs of initial parameters and only preserves the parameters with the highest AR after each round of optimization. The numerical simulations presented above suggest that MLI can get the quasi-optima by consuming fewer costs, besides, greedy-MLI gets a higher average performance.
	
	\medskip
	Reducing the training costs for parameterized quantum circuits in optimization tasks can accelerate the exploration of solution spaces and enhance the practicality of quantum optimization techniques. Under their efficiency, MLI and greedy-MLI might be subroutines or pre-processing tools in other optimization algorithms in future studies. Although our heuristic initialization strategies show advantages in cost and performance for all 3-regular graph instances we examine compared with the existing work, the performance of MLI and greedy-MLI on other graph ensembles or optimization tasks (e.g., portfolio optimization, logistics planning, and supply chain management) is worthwhile to explore. Finally, we hope that the M-Leap established in this work can inspire future research that could lead to a better understanding of what happens under the hood of QAOA optimization.

	% Acknowledgements
	\medskip
	\textbf{Acknowledgements} \par %delete if not applicable))
	We thank Xiang Guo, Shasha Wang and Yongmei Li for valuable suggestions. This work is supported by Beijing Natural Science Foundation (Grant No. 4222031), NSFC (Grant Nos. 62371069, 62272056, 62372048), and BUPT Excellent Ph.D. Students Foundation(CX2023123).
	
	\bibliography{refe}
	
	% Please provide Biographies and photos for Essays, Feature Articles, Progress Reports, Reviews, and Perspectives for those authors who should be highlighted  
	% These should be at most 100 words long
	% For other article types this section can be removed
	% Photographs should be 40mm broad and 50 mm high
	\clearpage
	\appendix
	\setcounter{figure}{0}
	\renewcommand{\thefigure}{A\arabic{figure}}
	\section{\textbf{The AR of MLI with various $l$ on a 3-regular graph with $n = 12$}}\label{AR_l_12}
\begin{figure*}[ht]
		\centering
		\setlength{\abovecaptionskip}{0.cm}
		\subfigure[The comparison of OAR]{
			\includegraphics[width=0.45\textwidth]{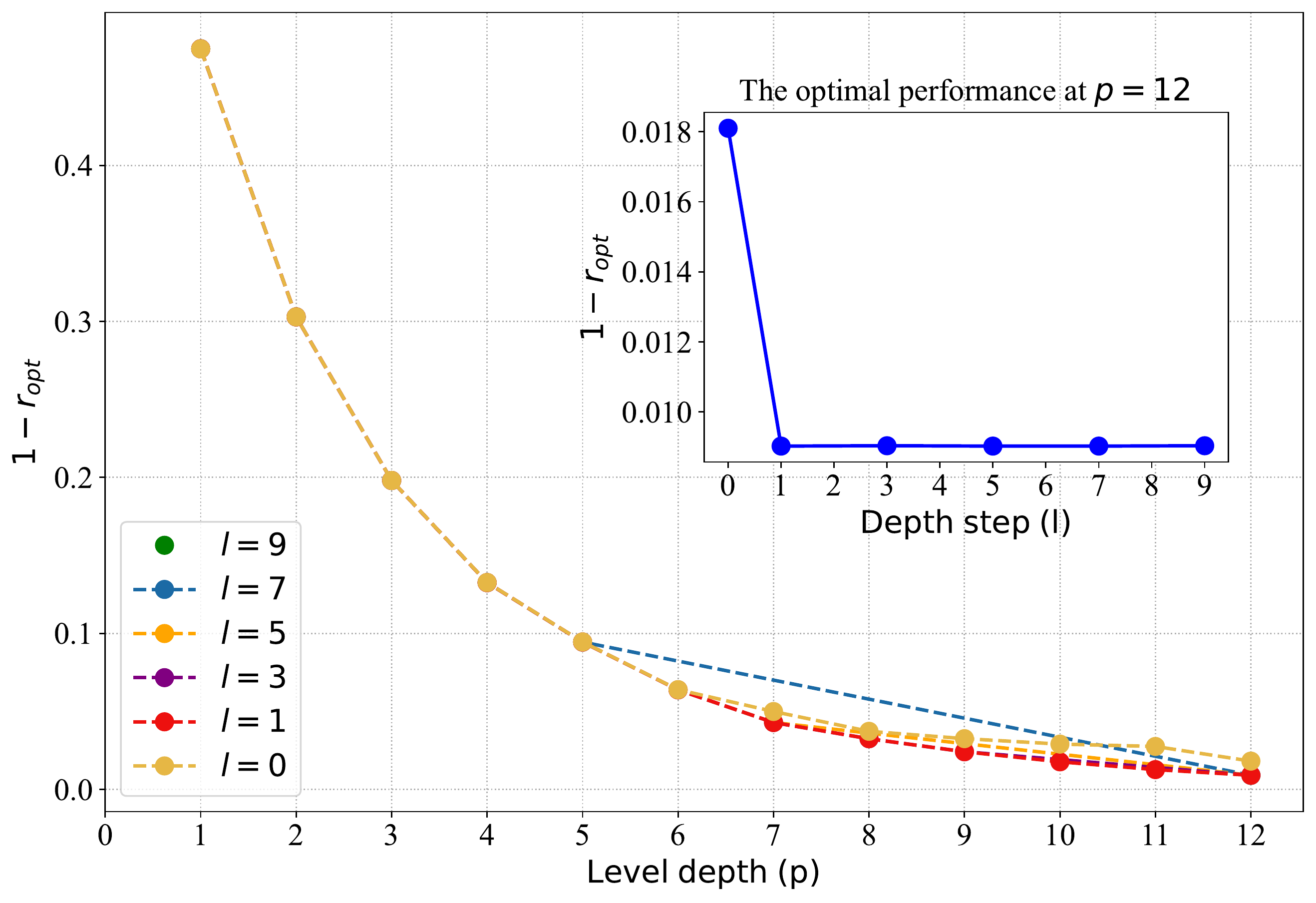}
		}
		\subfigure[The comparison of AAR]{
			\includegraphics[width=0.45\textwidth]{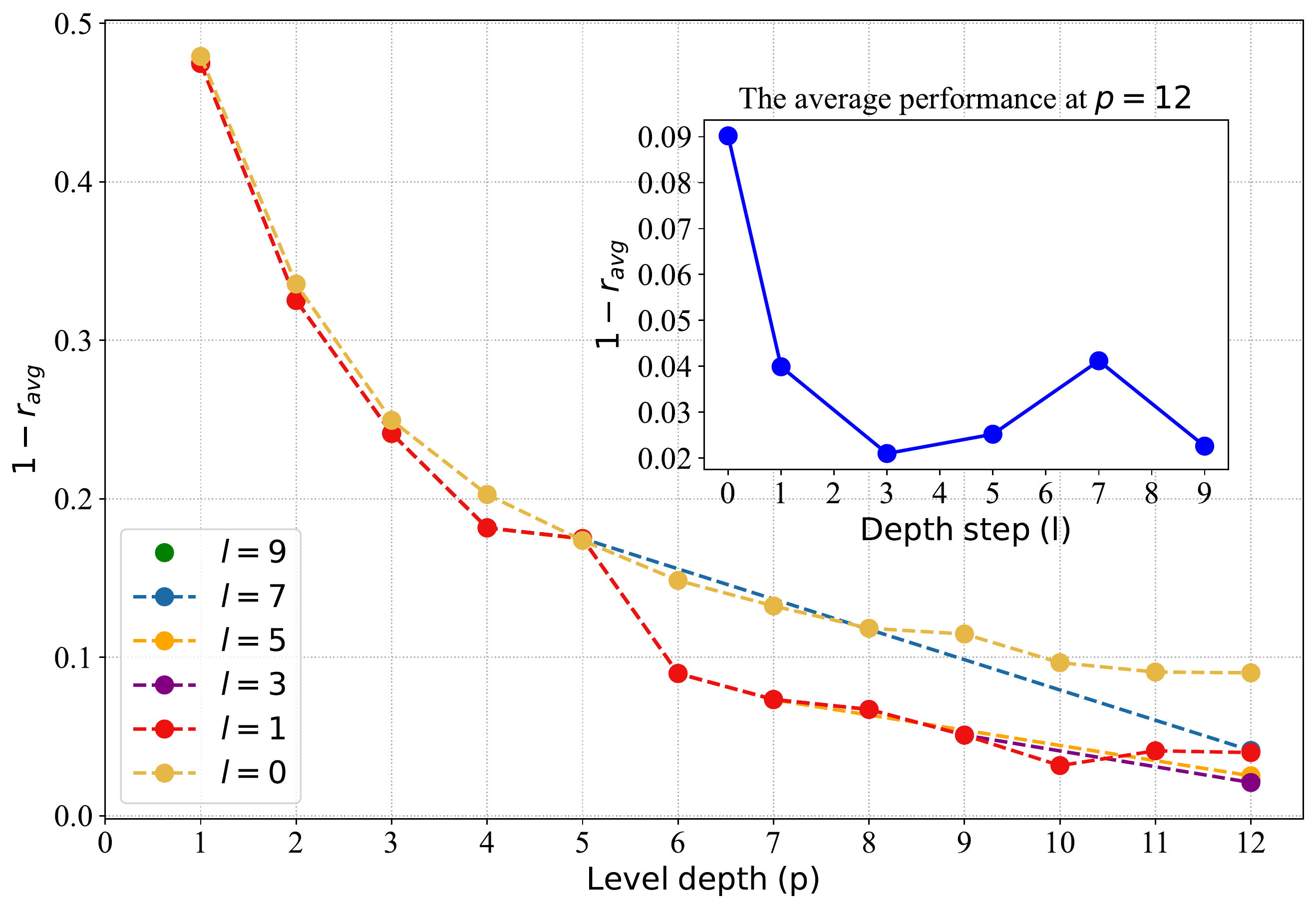}
		}
		\caption{The AR of MLI versus the level depth and the depth step $l$, where $l = 0$ and $l = 1$ respectively correspond to RI and INTERP. For each $l$, MLI achieves better performance as $p$ increases. (a) The OAR of MLI at each level depth. At shallow level depth, the quasi-optimal solution obtained by 500 runs of RI is the same as that obtained by INTERP (the yellow and red dots coincide completely), but with the increase of $p$, the quasi-optimal solution obtained by RI is lower than that obtained by INTERP. The inset depicts the OAR of MLI versus the depth step when $p=12$. We note that MLI with $l = 3,5,7,9$ can obtain the identically optimal performance as INTERP at the target level, and the OAR obtained by RI is slightly lower than the others when $p = 12$. (b) The AAR of MLI versus the level depth, where the green and purple dots are partly coincident when $p = 12$. The inset reflects the effects of $l$ on the AAR obtained by MLI at $p = 12$. These results show that the AAR obtained by RI is obviously lower than that of INTERP and MLI, and MLI can achieve a similar average performance to INTERP under $l = 7$ or a better AAR under $l = 3,5,9$.}
		\label{l_AR_12s}
	\end{figure*}

	We investigate the effect of $l$ on the performance of MLI on a 3-regular graph with $n = 12$. From the relevant experimental results in \textbf{Figure~\ref{l_AR_12s}}, we observe that MLI can obtain the same quasi-optimal solution as INTERP under certain $l$, and its AAR obtained at the target level depth is similar to INTERP but significantly higher than RI, as shown in Figure~\ref{l_ratio}. It is pointed out in Ref.~[34] that INTERP can obtain a quasi-optimal solution to the problem in polynomial optimization runs, while exponential optimization runs are required for RI to achieve similar performance. Besides obtaining a quasi-optimal solution to the problem in polynomial runs, MLI reduces the number of optimization rounds per run for INTERP and can obtain a higher AAR than RI. When there are more low-quality local solutions in the objective function landscape, RI is easy to fall into them. At this time, the performance benefits of INTERP and MLI become apparent as the scale of the problem increases.
	
	\section{\textbf{The calculation of running time}}\label{total_run_time}
	The time duration of a single repetition $t$ is given by the number of sequential quantum operations per quantum circuit. Ref.~\cite{runtime} considers it to be the sum of $t_{P}$, $t_{M}$, and the product of circuit depth and $t_{G}$, where $t_{P}$ is the time to prepare the initial state, $t_{G}$ is the average time duration of a quantum gate, and $t_{M}$ is the time per measurement. Same as in Ref.~\cite{runtime}, we use the same realistic if aggressive assumption of 1 microsecond for $(t_{P}+t_{M})$ and 10 nanoseconds for $t_{G}$ \cite{superconducting}. In this paper, we ignore the consumption of the time duration of the quantum circuit because the product of circuit depth and $t_{G}$ is too small. Thus, the time duration of a single repetition is about 1 microsecond.
	
	\medskip
	When the level depth is $p$, the running time required by INTERP in an optimization run is $\sum (t\times M \times T_{i, INT})$, where $i=1,\cdots,p$ and $T_{i,INT}$ represents the number of iterations consumed by INTERP at level depth $i$. By the same token, the time consumption of MLI in an optimization run is $\sum(t\times M \times T_{q_{j}, MLI})$, where $q_{j}$ represents the level depth in which MLI performs optimization. By analogy, the running time of greedy-MLI and RI are respectively $\sum_{c_{q_{j}}}\sum t\times M\times T_{q_{j, g}}$ and $t\times M\times T_{p, RI}$. Here, $c_{q_{j}}$ represents the sets of parameters at level depth $q_{j}$.

	\section{\textbf{The total iterations in multiple optimization rounds (runs)}}\label{total_ITR}
	
\begin{figure*}[htbp]
		\centering
		\setlength{\abovecaptionskip}{0.cm}
		\subfigure[$n = 8$]{
			\includegraphics[width=0.315\textwidth]{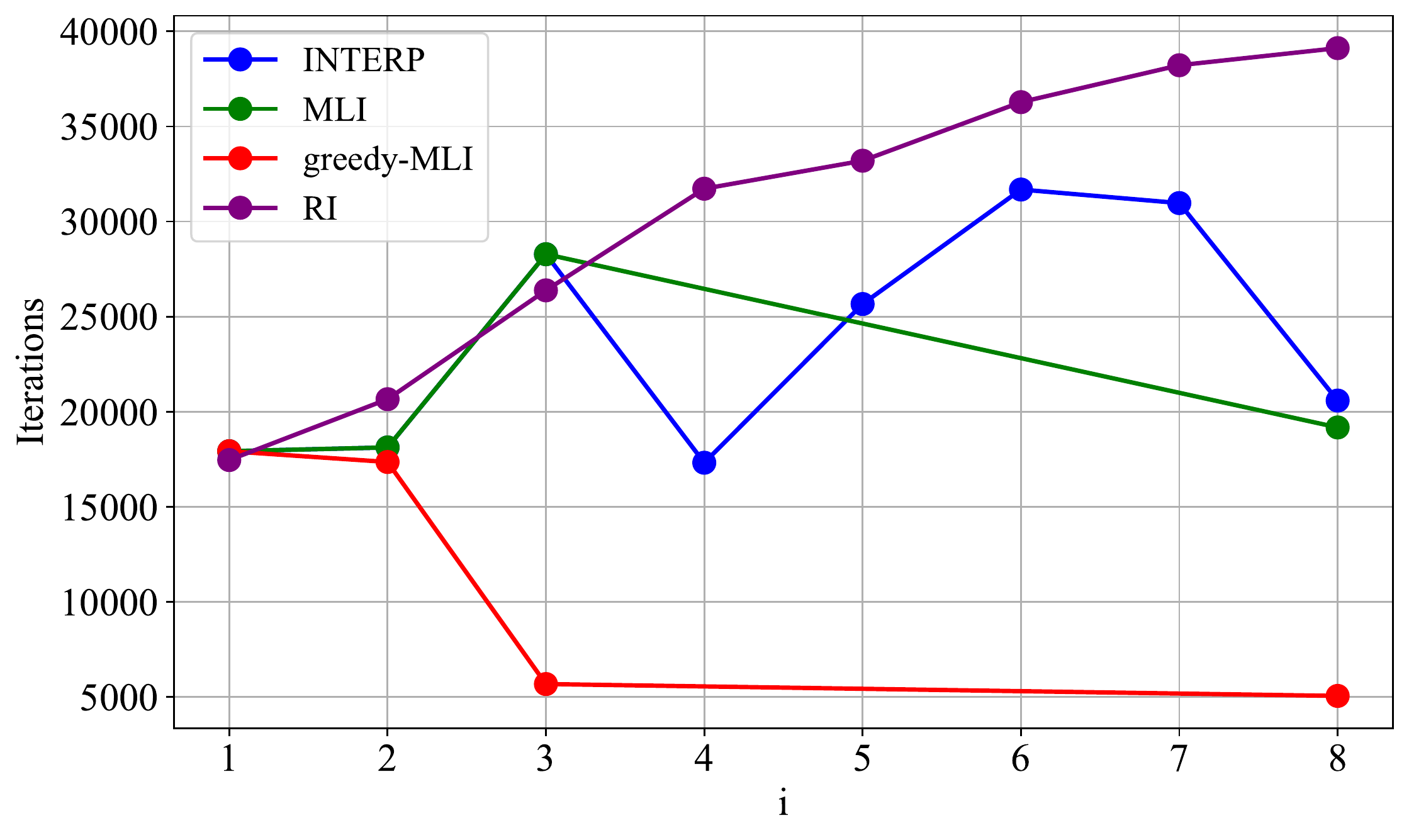}
		}
		\subfigure[$n = 10$]{
			\includegraphics[width=0.315\textwidth]{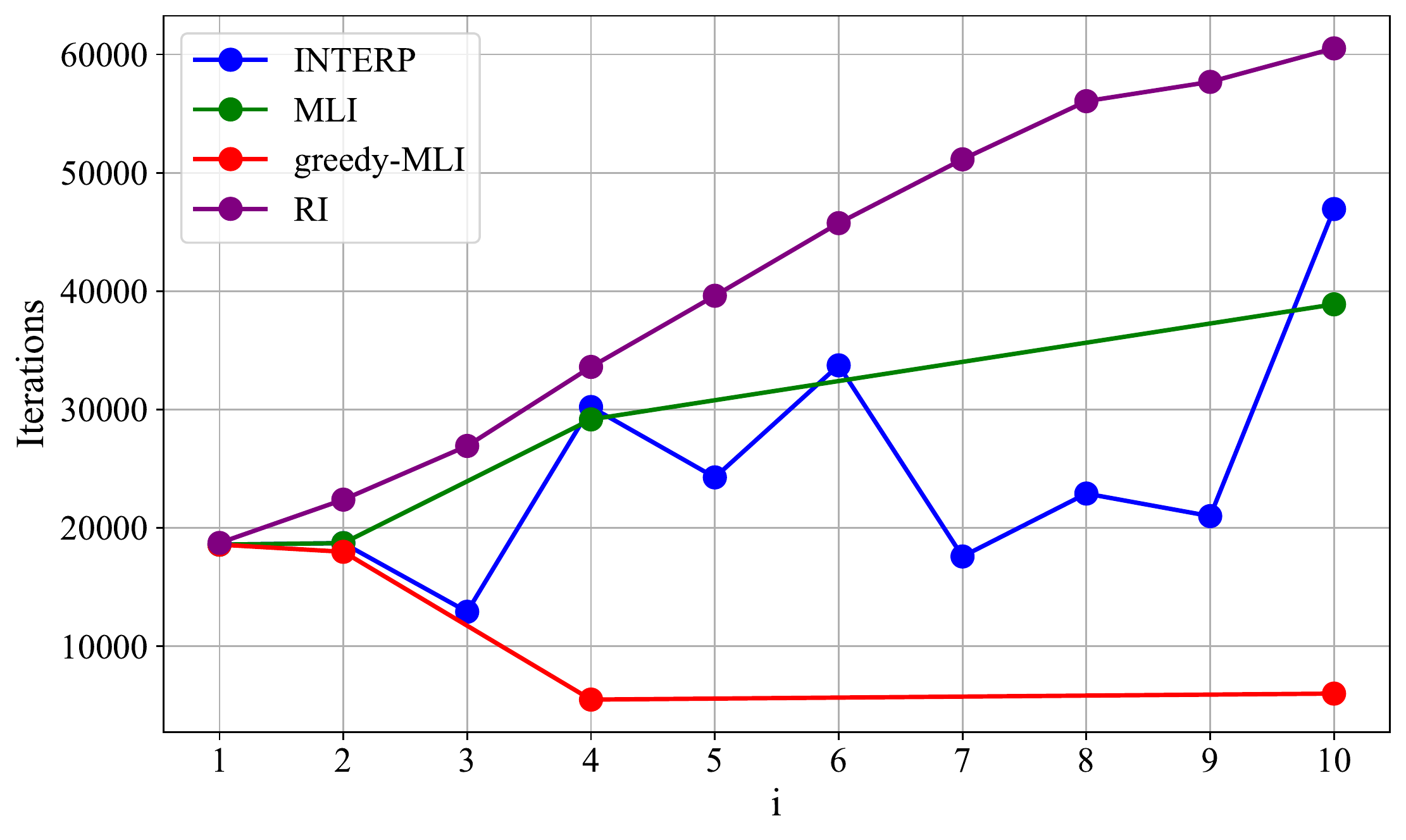}
		}
		\subfigure[$n = 12$]{
			\includegraphics[width=0.315\textwidth]{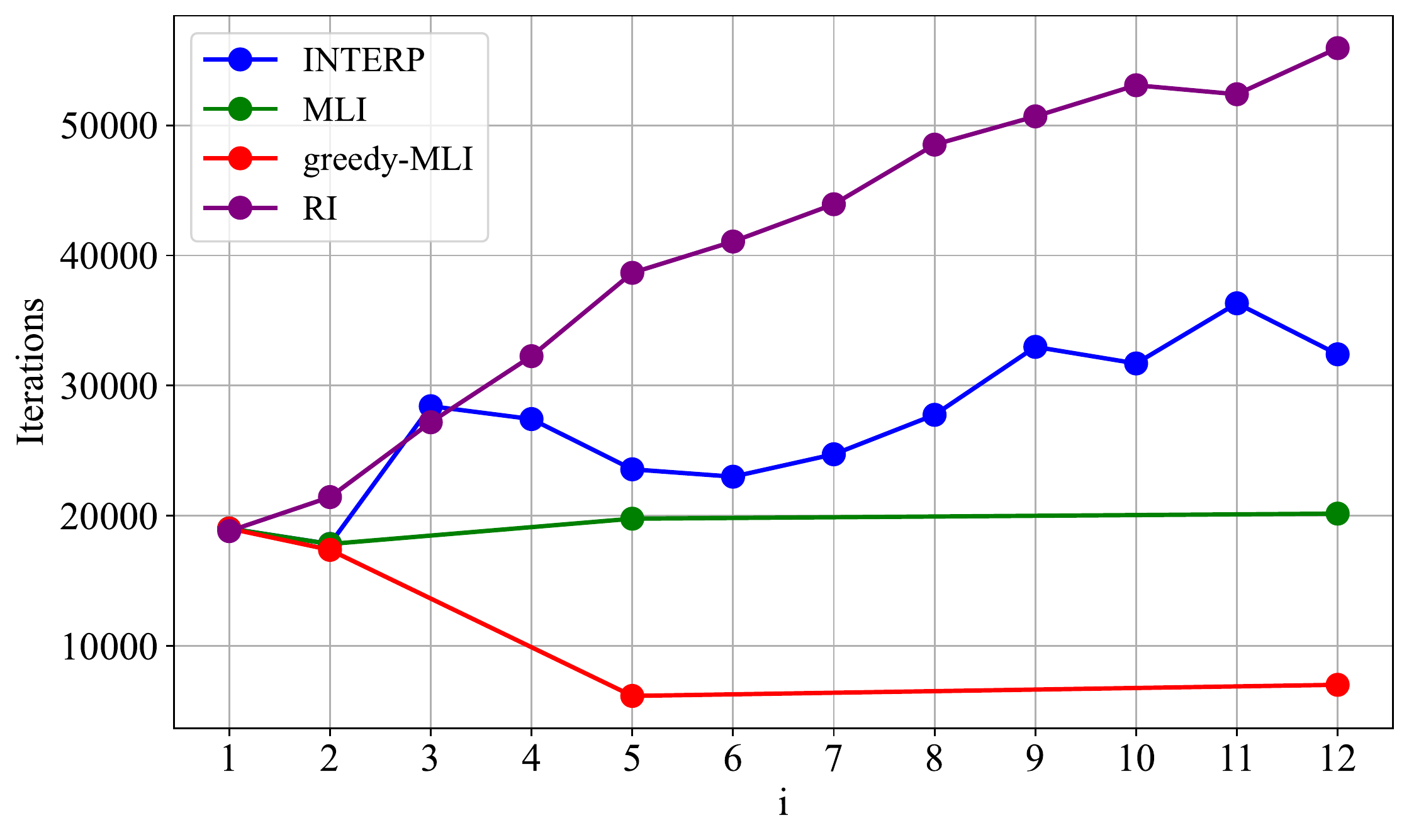}
		}
		\caption{The total number of consumed iterations in multiple optimization rounds for various level depths $i$. Different sub-figures correspond to various graph instances. For INTERP, MLI, and greedy-MLI, the accumulation of these points on the same curve is the total number of iterations in 500 (or an) optimization runs, as shown in Figure~\ref{ITR_vs} in this appendix. For RI, the total number of iterations in 500 optimization runs at $p$ equals the consumption in 500 optimization rounds. From these results, we can observe that the consumption of RI dramatically increases as $i$ increases, which is opposite to the greedy-MLI. The low consumption of greedy-MLI benefits from the operation of discarding non-ideal parameters, thus improving the quality of parameters and decreasing the number of optimization rounds compared with other strategies. }
		\label{total_iterations}
	\end{figure*}

\begin{figure*}[htbp]
		\centering
		\setlength{\abovecaptionskip}{0.cm}
		
		\subfigure[The total iterations  when $n = 8$]{
			\includegraphics[width=0.315\textwidth]{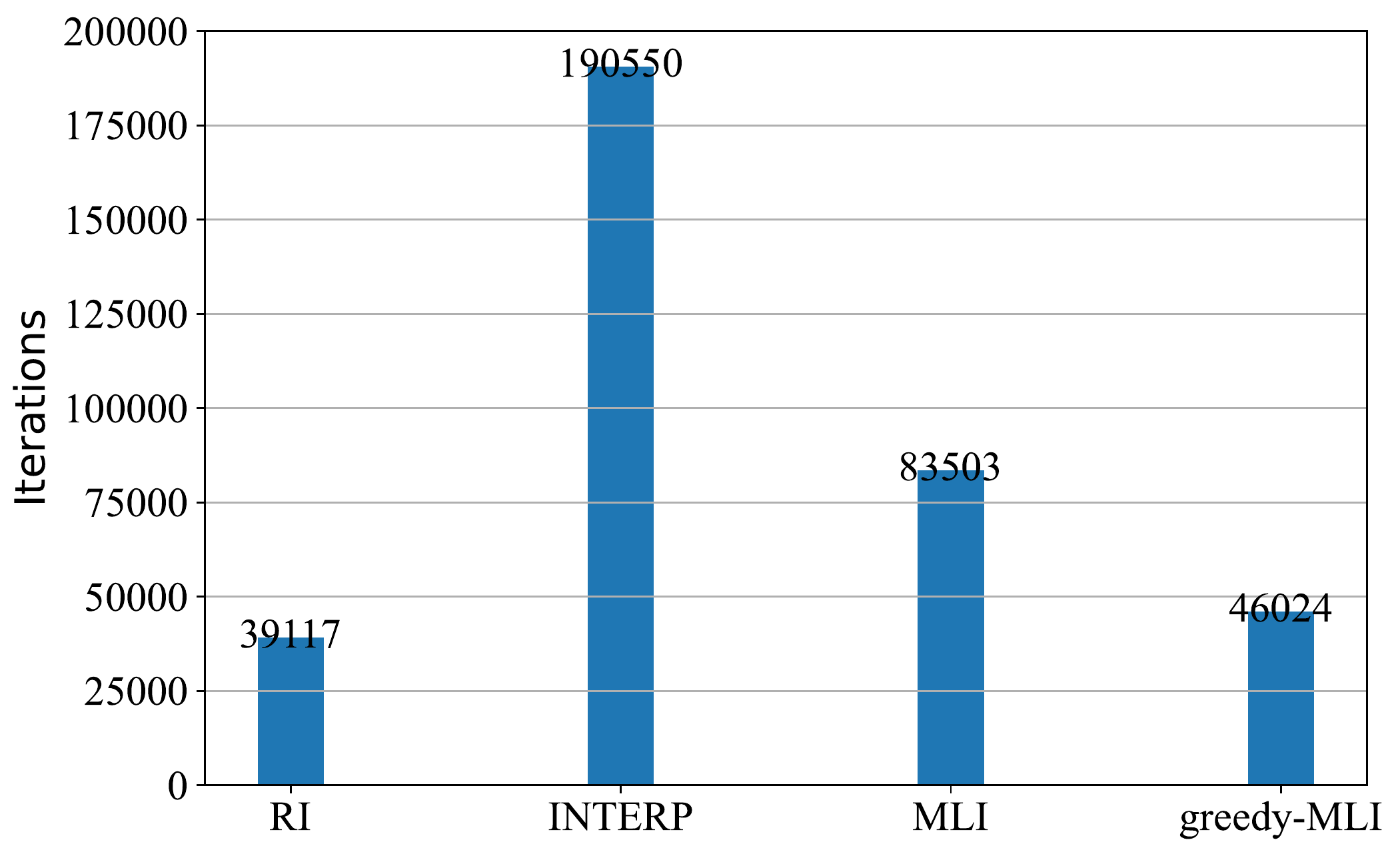}
			\label{ITR_vs:subfig1}
		}
		\subfigure[The total iterations when $n = 10$]{
			\includegraphics[width=0.315\textwidth]{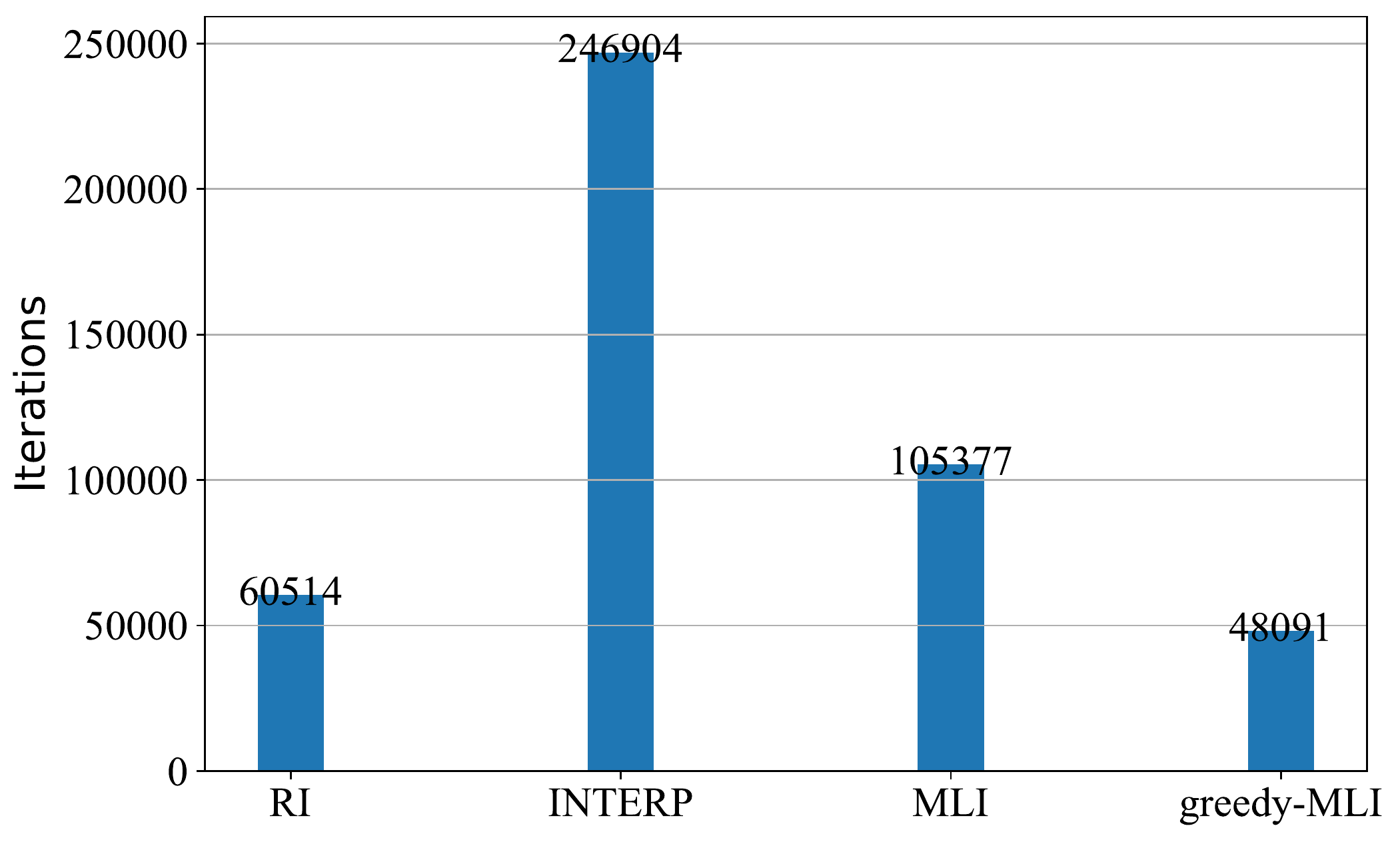}
			\label{ITR_vs:subfig2}
		}
		\subfigure[The total iterations when $n = 12$]{
			\includegraphics[width=0.315\textwidth]{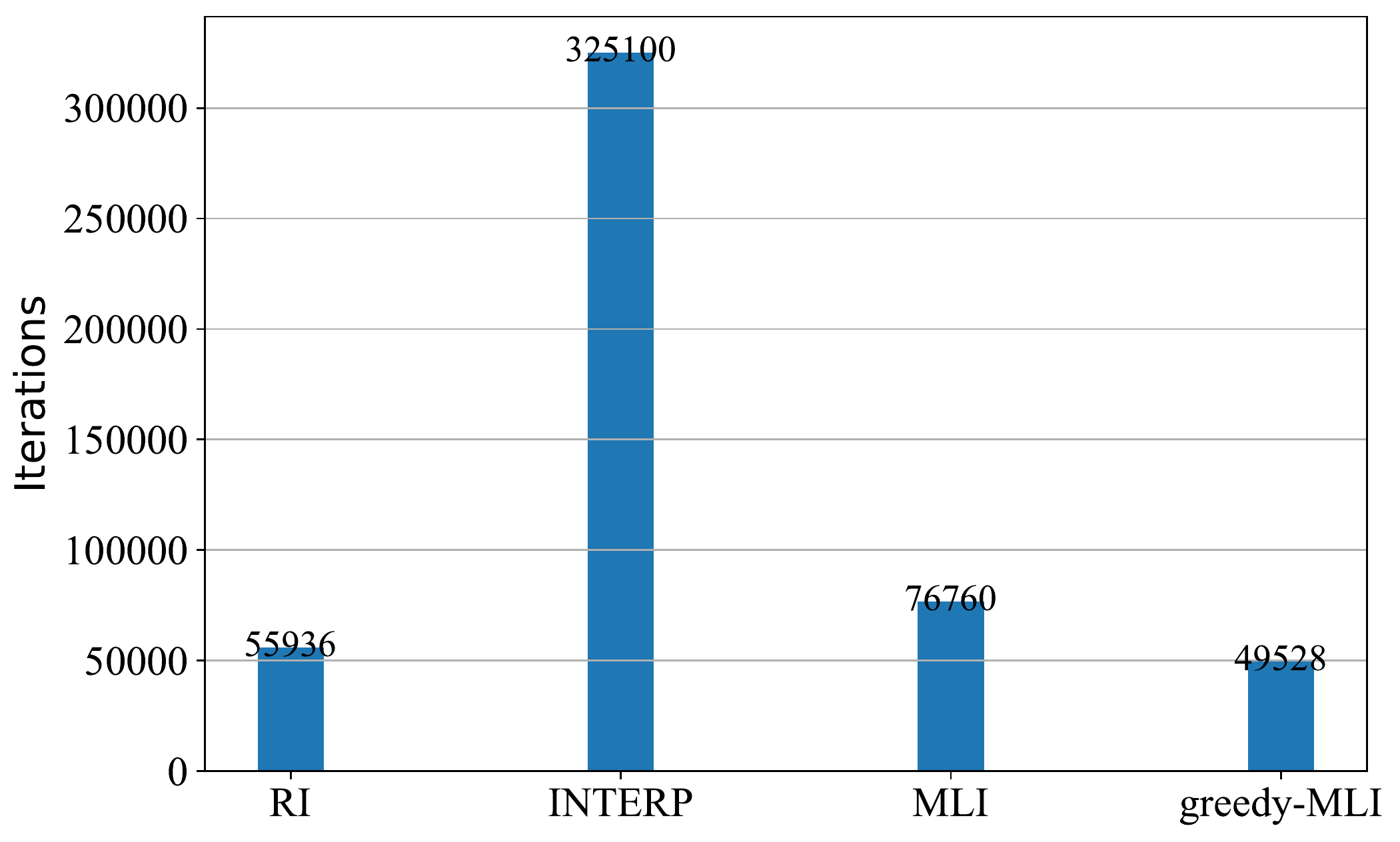}
			\label{ITR_vs:subfig3}
		}
		\caption{The iteration comparison of four strategies on 3-regular graphs with vertex numbers $n=8,10,12$. Here, we respectively calculate the total number of iterations for 500 optimization runs of RI, INTERP, MLI, and one optimization run of greedy-MLI. These results show that the total number of iterations consumed by MLI and greedy-MLI is approximately 1/2 and 1/5 of that consumed by INTERP.}\label{ITR_vs}
	\end{figure*}

	In our simulations, we perform 500 optimization runs for RI, INTERP, and MLI, and an optimization run of greedy-MLI with $c = 500$ on each graph instance. For each level depth where INTERP (RI, MLI, and greedy-MLI) executes optimization, we can get 500 (500, 500, $c_{q_{j}}$) data points that respectively correspond to the consumed iterations in various optimization rounds, where $c_{q_{j}}$ represents the sets of parameters at level depth $q_{j}$. For each strategy, we can get the total number of iterations consumed in multiple optimization rounds at the current level depth by accumulating their corresponding data points. The detailed number of iterations in multiple optimization rounds for various level depths is given in \textbf{Figure~\ref{total_iterations}}. The total number of iterations in multiple optimization runs is shown in \textbf{Figure~\ref{ITR_vs}}.
	
\end{document}